\documentclass[epjST]{svjour}
\usepackage{graphics,amsmath,amssymb,amsfonts,bbm}

\def\bmA{\boldsymbol{A}}
\def\bmB{\boldsymbol{B}}
\def\bmC{\boldsymbol{C}}

\def\bmM{\boldsymbol{M}}
\def\bmR{\boldsymbol{R}}
\def\bmS{\boldsymbol{S}}
\def\bmW{\boldsymbol{W}}
\def\bmX{\boldsymbol{X}}

\def\bmdelta{\boldsymbol{\delta}}
\def\bmDelta{\boldsymbol{\Delta}}
\def\bmsigma{\boldsymbol{\sigma}}
\def\bmSigma{\boldsymbol{\Sigma}}
\def\bmLambda{\boldsymbol{\Lambda}}
\def\bmOmega{\boldsymbol{\Omega}}
\def\sT{{\scriptscriptstyle T}}
\def\rr{\mathbb{R}}

\def\cc{\mathbb{C}}
\def\ii{{\mathbbm 1}}

\def\rma{{\rm a}}
\def\rmb{{\rm b}}

\def\hR{\hat{R}}

\def\ha{\hat{a}}

\def\hb{\hat{b}}

\def\hO{\hat{O}}
\def\Tr{\hbox{Tr}}
\def\mL{{\cal L}}
\def\mD{{\cal D}}
\def\bmzero{\boldsymbol{0}}
\newcommand{\vset}{{\boldsymbol v}}
\newcommand{\ket}[1]{| #1 \rangle }

\newcommand{\eq}[1]{Eq.~(\ref{#1})}
\newcommand{\gr}[1]{\boldsymbol{#1}}
\newcommand{\be}{\begin{equation}}
\newcommand{\ee}{\end{equation}}
\newcommand{\bea}{\begin{eqnarray}}
\newcommand{\eea}{\end{eqnarray}}
\newcommand{\mbfrac}{\mbox{$\frac12$}}

\newcommand{\refeq}[1]{Eq.~(\ref{#1})}
\def\re#1{\Re\hbox{e}[#1]}
\def\im#1{\Im\hbox{m}[#1]}
\begin{document}
\title{Quantum optics in the phase space}
\subtitle{A tutorial on Gaussian states}
\author{Stefano Olivares\inst{1,2}
\fnmsep\thanks{\email{stefano.olivares@ts.infn.it}}}
\institute{Dipartimento di Fisica, Universit\`a degli Studi di
  Trieste, I-34151 Trieste, Italy \and CNISM UdR Milano Statale,
  I-20133 Milano, Italy}
\abstract{In this tutorial, we introduce the basic concepts and
  mathematical tools needed for phase-space description of a very
  common class of states, whose phase properties are described by
  Gaussian Wigner functions: the Gaussian states. In particular, we
  address their manipulation, evolution and characterization in view
  of their application to quantum information.} %end of abstract
\maketitle
%
%%%%%%%%%%%%%%%%%%
%\newpage
\tableofcontents
%%%%%%%%%%%%%%%%%%%%%%%% SECTION %%%%%%%%%%%%%%%%%%%%%%%%%%
\section{Introduction}\label{intro}
A Gaussian state is a state with Gaussian Wigner functions
\cite{schu:86}. In recent years, an increasing attention has been
devoted to this class of states, as they play a major role in quantum
information with continuous variables
\cite{FOP:05,bra:05,wee:11}. Besides quantum optics, where they are
generated with current quantum optical technology, Gaussian states
naturally appear in the description of optomechanical and
nanomechanical oscillators, gases of cold atoms and ion
traps. Furthermore, Gaussian states exhibit extremality properties:
among the continuous variable states, they tend to be extremal if one
imposes some constraints on the covariance matrix \cite{wolf:06}.
\par
The better way to deal theoretically with Gaussian states is to use a
suitable phase-space analysis. In fact, if the Gaussian character is
preserved during the dynamics, it is natural to think of the evolution
of a Gaussian state as a transformation of the covariance matrix and
first-moments vector that fully characterize it. As we will see, this
is the case if we consider the linear and bilinear interactions used
in quantum optics labs to generate and manipulate quantum
states. Furthermore, the Gaussian properties of these states may be
preserved also during the dissipative evolution through noisy
channels, both Markovian and non-Markovian.
\par
The main purpose of this tutorial is to introduce the reader to the
phase-space description of Gaussian states in view of their
applications to quantum information. After the definition of Gaussian
state and its basic properties in relation with the covariance matrix
and first-moments vector (Sect.~\ref{s:notation}), we will focus on
the unitary evolution through suitable symplectic transformations
(Sect.~\ref{s:symp:trasf}). We will illustrate the phase-space
approach to describe the generation and manipulation of Gaussian
states with linear and bilinear interaction of modes
(Sect.~\ref{s:lin:bilin}). Single-mode Gaussian states will be briefly
addressed in Sect.~\ref{sec:sm:GS}, while we will discuss two-mode
Gaussian states in more details in Sect.~\ref{sec:tm:GS}. In the
latter case, the concepts of symplectic eigenvalues as well as of
standard form of the covariance matrix and local symplectic invariants
will be introduced and applied to calculate the entropies and the
mutual information of two-mode Gaussian states and to investigate
their separability, entanglement and Gaussian quantum discord
(Sect.~\ref{sec:tm:GS}). Dynamics through Markovian noisy channels
will be addressed in Sect.~\ref{sec:GS:noisy} and
Sect.~\ref{sec:G:meas} will consider the effect of a Gaussian
measurement performed on a multimode Gaussian state. We will review
the main results concerning the fidelity between Gaussian states in
Sect.~\ref{sec:fidelity}. Sect.~\ref{sec:remarks} will close the tutorial
with some concluding remarks.
%
%%%%%%%%%%%%%%%%%%%%%%%% SECTION %%%%%%%%%%%%%%%%%%%%%%%%%%
\section{Basic notation and definition of Gaussian state}
\label{s:notation}
Each mode $k=1,\ldots, n$ of a system made of $n$ modes or, more in
general, $n$ bosons is described by the annihilation and creation
operators $\hat{a}_k$ and $\hat{a}_k^{\dag}$, respectively, with
commutation relations $[\hat{a}_k,\hat{a}^\dag_l]=\delta_{kl}$. The
Hilbert space of the whole system, ${\cal H}=\otimes_{k=1}^n \: {\cal
  F}_{k}$, is the tensor product of the infinite dimensional Fock
spaces ${\cal F}_{k}$ of the $n$ modes, each spanned by the number
basis $\{\ket{m}_k\}_{m \in {\mathbb N}}$, i.e., by the eigenstates of
the number operator $\hat{a}_k^{\dag}\hat{a}_k$. If we address
non-interacting modes and use the natural units, the free Hamiltonian
of the system may be simply written as $H=\sum_{k=1}^n
(\ha_k^{\dag}\ha_k + \frac12)$ and the corresponding position- and
momentum-like operators for the $k$-th mode are:
\begin{eqnarray}
\hat{q}_k = \frac{1}{\sqrt{2}}\,(\hat{a}_k + \hat{a}^\dag_k)\,, \quad
\hbox{and} \quad
\hat{p}_k = \frac{1}{i\sqrt{2}}\,(\hat{a}_k - \hat{a}_k^\dag)\,,
\label{defQP}\;
\end{eqnarray}
respectively.
The commutation relations $[\hat{q}_k,\hat{p}_l]=i\, \delta_{kl}$
associated with $\hat{q}_k$ and $\hat{p}_l$ can be rewritten in the
following compact form, which will turn out to be very useful for the
phase-space analysis:
\begin{eqnarray}
\left[\hat{R}_k,\hat{R}_l\right]=i\, \Omega_{kl}
\label{defCom1}\;,
\end{eqnarray}
where $\hat{\boldsymbol{R}}=(\hat{q}_1,\hat{p}_1,\ldots,
\hat{q}_n,\hat{p}_n)^{\sT}$ is a vector of operators and $\Omega_{kl}
\equiv [\gr{\Omega}]_{kl}$ are the elements of the symplectic matrix:
\begin{equation}
\gr{\Omega}=\bigoplus_{k=1}^{n}\gr{\omega}\:, \qquad \gr{\omega}=
\left(\begin{array}{cc}0&1\\ -1&0
\end{array}\right)\,. \label{defOM}
\end{equation}
Note that $\bmOmega^\sT = -\bmOmega = \bmOmega^{-1}$.
\par
We can now introduce the leading element of this tutorial. A $n$-mode
state described by the density matrix $\varrho$ is a Gaussian state if
its characteristic function:
\begin{equation}\label{CHI:def}
\chi[\varrho](\bmLambda) =
\mbox{Tr}\left[\varrho\: \exp\left\{
-i \bmLambda^{\sT}\bmOmega \hat{\bmR}
\right\}\right]
\end{equation}
is Gaussian, namely, if $\chi[\varrho](\bmLambda)$ can be written in the
following form:
\begin{equation}\label{CHI:GS}
\chi[\varrho](\bmLambda) =\exp\left\{
-\mbfrac\bmLambda^{\sT} \bmOmega \bmsigma \bmOmega^{\sT} \bmLambda
-i \bmLambda^{\sT} \bmOmega \langle\hat{\bmR}\rangle
\right\}\,,
\end{equation}
with $\boldsymbol{\Lambda} =
(\rma_1,\rmb_1,\ldots,\rma_n,\rmb_n)^{\sT} \in \rr^{2n}$ and we
defined the covariance matrix (CM):
\begin{align}\label{defCOV}
\sigma_{kl} \equiv [\bmsigma]_{kl} =
\mbfrac \langle \hR_k,\hR_l + \hR_l,\hR_k \rangle - 
\langle  \hR_k \rangle  
\langle  \hR_l \rangle\,,
\end{align}
and $\langle \hat{O} \rangle = \hbox{Tr}[\varrho\: \hat{O}]$ is the
expectation value of the operator $\hat{O}$. The vector
$\langle\hat{\bmR}\rangle \equiv \hbox{Tr}[\varrho\: \hat{\bmR}]$ is
usually referred to as first-moments vector. The uncertainty relations
among canonical operators impose a constraint on the CM, corresponding
to the inequality \cite{simon_old1,simon_old2}:
\begin{equation}\label{HeisSG}
\boldsymbol{\sigma} + \frac{i}{2}\, \boldsymbol{\Omega} \geq 0\,,
\end{equation}
that expresses, in a compact form, the positivity of the density matrix
$\varrho$.
\par
The exponential appearing in Eq.~(\ref{CHI:def}) is called
displacement operator:
\begin{eqnarray}
  \exp\left\{-i \boldsymbol{\Lambda} ^{\sT}\boldsymbol{\Omega}
\hat{\boldsymbol{R}} \right\} = D(\boldsymbol{\Lambda}) \equiv
  D(\boldsymbol{\lambda}) = \bigotimes_{k=1}^n D_k(\lambda_k)\,, 
\label{defDcmplx}\;
\end{eqnarray}
where $\boldsymbol{\lambda} = (\lambda_1,\ldots,\lambda_n)^{\sT} \in
\cc^n$, $\lambda_k = \frac{1}{\sqrt{2}} (\rma_k + i \rmb_k)$, and
$D_k(\lambda_k)=\exp \{\lambda_k \hat{a}^\dag_k - \lambda_k^*
\hat{a}_k \}$ are single-mode displacement operators acting on the
$k$-th mode. Displacement operator takes its name after its
action on the mode operators, namely:
\begin{align}
D^\dag (\boldsymbol{\lambda})\, \hat{a}_k\, D(\boldsymbol{\lambda})
= \hat{a}_k + \lambda_k, \quad \hbox{and \quad}
D^\dag (\boldsymbol{\Lambda})\, \hat{\boldsymbol{R}}\, D(\boldsymbol{\Lambda}) = \hat{\boldsymbol{R}} + \boldsymbol{\Lambda}\,.
\label{displa_mode}\;
\end{align}
In other words, in phase space composed of the couples of conjugate
variables $\hat{q}_k$ and $\hat{p}_k$, $k=1,\ldots, n$, it displaces a
state by an amount $\boldsymbol{\Lambda}$ \cite{gla:63}.
\par
By Fourier transforming the characteristic function (\ref{CHI:def}),
we obtain the so-called Wigner function of $\varrho$
\cite{cah:69,wig:32}:
\begin{equation}
W[\varrho](\gr{X}) = \frac{1}{(2\pi^{2})^{n}}
\int_{\rr^{2n}}
d^{2n} \boldsymbol{\Lambda} \:
\exp\left\{ i \boldsymbol{\Lambda}^{\sT}\bmOmega
\boldsymbol{X}\right\} \chi[\varrho] (\gr{\Lambda})
 \label{defWCarta}\,,
\end{equation}
with $\boldsymbol{X} = (x_1,y_1,\ldots,x_n,y_n)^{\sT}$. Note that:
\begin{equation}\label{delta:1}
\frac{1}{\pi^{2n}}\int_{\rr^{2n}}
d^{2n} \boldsymbol{\Lambda} \:
\exp\left\{ i \boldsymbol{\Lambda}^{\sT}\bmOmega
\boldsymbol{X}\right\} = 2^{n}\, \delta^{(2n)}(\bmX),
\end{equation}
$\delta^{(2n)}(\bmX)$ being the $2n$-dimensional $\delta$-function.
From the identity:
\begin{equation}
\int_{\rr^{n}}d^{2n}\bmLambda \exp\left\{
-\hbox{$\frac12$}\bmLambda^{\sT} \boldsymbol{Q} \bmLambda
+i \bmLambda^{\sT}\bmX \right\} =
\frac{(2\pi)^n
\exp\left\{-\hbox{$\frac12$}\bmX^{\sT} \boldsymbol{Q}^{-1} \bmX
\right\}}{\sqrt{\det[\boldsymbol{Q}]}}\,,
\end{equation}
where $\boldsymbol{Q}$ is a real, positive-definite symmetric $2n
\times 2n$ matrix, it follows that in the case of the Gaussian state
(\ref{CHI:GS}) we have:
\begin{equation}\label{c3:WX}
W[\varrho](\bmX) =
\frac{\exp\left\{ -\frac12
\left(\bmX-\langle\hat{\bmR}\rangle\right)^{\sT} \bmsigma^{-1}
\left(\bmX-\langle\hat{\bmR}\rangle\right) \right\}}
{\pi^n\,\sqrt{\det[\bmsigma]}}\,,
\end{equation}
that is still Gaussian. It is possible to show that pure Gaussian
states are the only pure states with positive Wigner function
\cite{Hud74,LB95}.
\par
The same Wigner function (\ref{c3:WX}) can be
calculated also as follows:
\begin{equation}
W[\varrho](\gr{X}) = \frac{1}{(2\pi^{2})^{n}}
\int_{\rr^{2n}}
d^{2n} \boldsymbol{\Lambda} \:
\exp\left\{ i \boldsymbol{\Lambda}^{\sT}
\boldsymbol{X}\right\} \tilde{\chi}[\varrho] (\gr{\Lambda})
 \label{defWCartb}\,,
\end{equation}
where $\tilde{\chi}[\varrho] (\gr{\Lambda}) =
\chi[\varrho] (\bmOmega\gr{\Lambda})$, namely:
\begin{equation}\label{chi:simple}
\tilde{\chi}[\varrho] (\gr{\Lambda}) =
\exp\left\{
-\mbfrac\bmLambda^{\sT} \bmsigma \bmLambda
-i \bmLambda^{\sT} \langle\hat{\bmR}\rangle
\right\}\,.
\end{equation}
The equivalence between Eq.~(\ref{defWCarta}) and Eq.~(\ref{defWCartb})
is due to the fact that an equivalent definition of
Eq.~(\ref{delta:1}) is:
\begin{equation}
\frac{1}{\pi^{2n}}\int_{\rr^{2n}}
d^{2n} \boldsymbol{\Lambda} \:
\exp\left\{ i \boldsymbol{\Lambda}^{\sT}
\boldsymbol{X}\right\} = 2^{n}\, \delta^{(2n)}(\bmX).
\end{equation}
Eq.~(\ref{c3:WX}) is a particular case of the
more general $s$-ordered Wigner function \cite{cah:69,MF:97}:
\begin{equation}
W_s[\varrho](\gr{X}) =
\frac{1}{(2\pi^2)^{n}} \int_{\rr^{2n}}
d^{2n}\!\gr{\Lambda}\:
\exp\left\{\mbox{$\frac{1}{2}$} s |\bmLambda|^2
+i \bmLambda^{\sT}\bmOmega \bmX \right\} \: \chi[\varrho](\gr{\Lambda}).
\label{desWs}
\end{equation}
If $s=0$, then we have the usual Wigner function (\ref{c3:WX}); if
$s=-1$ or $s=1$ we obtain the Husimi $Q$-function or the
Glauber-Sudarshan $P$-function, respectively
\cite{cah:69}. Furthermore, by using the relation:
\begin{equation}
W_s[\varrho](\bmX) = \int_{\rr^{2n}}
d^{2n}\gr{Y}\: \frac{1}{\pi(r-s)}
\exp\left\{  -\frac{|\gr{Y}-\gr{X}|^2}{r-s} \right\}\,
 W_r[\varrho](\gr{Y}),
\end{equation}
a $r$-ordered Wigner function can be transformed into a $s$-ordered
one. The $s$-ordered Wigner function is used to define the nonclassical
depth ${\cal T}$ of a quantum state \cite{lee:91}:
\begin{equation}
{\cal T} = \frac12 (1-\overline{s}),
\end{equation}
where $\overline{s}$ is the maximum value for which
$W_s[\varrho](\gr{X})$ becomes positive and semidefinite, i.e., a
probability distribution. One has ${\cal T} = 1$
for number states and ${\cal T} = 0$ for coherent states.  The
nonclassical depth can be interpreted as the minimum number of thermal
photons that has to be added to a quantum state in order to erase all
the quantum features of state \cite{lee:91}.
\par
A Gaussian state is fully characterized by its CM and first-moments
vector. For instance, the purity $\mu(\varrho) = \mbox{Tr}[\varrho^2]$ of the
Gaussian state depends only on its CM and reads:
\begin{equation}\label{c3:purity:gauss:N}
\mu(\varrho) =
\frac{1}{2^{n}\,\sqrt{\mbox{Det}[\bmsigma]}}\,,
\end{equation}
where we used the trace rule in the phase space:
\begin{equation}\label{Wtrace}
\hbox{Tr}\left[\hO_1\: \hO_2 \right] = \left(\frac{\pi}{2}\right)^{n}
\int_{R^{2n}}d^{2n}\!\bmX\: W[\hO_1] (\bmX) \: W[\hO_2]
(\bmX)\:,
\end{equation}
which follows from the expansion:
\begin{equation}\label{O:expansion}
\hO = \int_{R^{2n}}d^{2n}\!\bmX\: W[\hO] (\bmX)\:
 D(\bmX) \boldsymbol{\Pi} D^{\dag}(\bmX)\,,
\end{equation}
where $\boldsymbol{\Pi} = \otimes_{k=1}^{n}(-1)^{\ha^{\dag}\ha}$ is
the parity operator and $ D(\bmX) \boldsymbol{\Pi} D^{\dag}(\bmX) =
D(2\bmX) \boldsymbol{\Pi} =  \boldsymbol{\Pi} D^{\dag}(2\bmX)$,
or, equivalently, by using the characteristic function formalism:
\begin{equation}\label{Chi:trace}
\hbox{Tr}\left[\hO_1\: \hO_2 \right] = \frac{1}{(2\pi)^{n}}
\int_{R^{2n}}
d^{2n}{\bmLambda}\: \chi[\hO_1] (\bmLambda) \:
\chi[\hO_2] (-\bmLambda)\:,
\end{equation}
which follows from:
\begin{equation}
\hO = \frac{1}{(2\pi)^{n}}\int_{R^{2n}}d^{2n}\!\bmLambda\:
\chi[\hO] (\bmLambda)\: D^{\dag}(\bmLambda)\,.
\end{equation}
We recall also that $\hbox{Tr}[D(\bmLambda)] = (2\pi)^n
\delta^{(2n)}(\bmLambda)$ and $\hbox{Tr}[D(\bmX)] = (2\pi)^n
\delta^{(2n)}(\bmX)$. Starting from Eq.~(\ref{O:expansion}) we can
also obtain the trace form for the Wigner function:
\begin{equation}
  W[\hO] = \left(\frac{2}{\pi}\right)^n
\hbox{Tr}[\hO\: D(\bmX) \boldsymbol{\Pi} D^{\dag}(\bmX)].
\end{equation}
\par
Note that the identity operator for $n$ modes has a Wigner function
given by $W[{\mathbbm I}] (\gr{X})= \pi^{-n}$, thus, form
Eq.~(\ref{Wtrace}) we have $\hbox{Tr}[\hO]=2^{-n}\int_{\cc^n}
d^{2n}\gr{X} \: W[\hO](\gr{X})$, from which follows the normalization of
the Wigner function (\ref{c3:WX}). The Wigner function formalism allows
to easily calculate the expectations of symmetrically ordered products
of field operators \cite{weyl:50}, namely:
\begin{equation}\label{exp:W}
\hbox{Tr}\left[\varrho\left[(\ha_s^{\dag})^{h}\,\ha_t^{k}\right]_{\rm s}\right] =
\frac{1}{2^{n}}\int_{\rr^{2n}} d^{2n}\bmX\,
W[\varrho](\bmX)\, (\alpha_s^*)^{h}\,\alpha_t^{k}\,,
\end{equation}
with, as usual, $\bmX=(x_1,y_1,\ldots,x_n,y_n)^{\sT}$,
$\alpha_k=\frac{1}{\sqrt{2}}(x_k+iy_k)$, and:
\begin{equation}
\left[(\ha_s^{\dag})^{h}\,\ha_t^{k}\right]_{\rm s} =
\left. \frac{\partial^{h+k}}{\partial x^h\,\partial y^k} 
\frac{(x\, \ha_s^{\dag} + y\, \ha_t)^{k+h}}{(k+h)!}\right|_{x=y=0}\,.
\end{equation}
For the sake of completeness, we observe that the expectations in
Eq.~(\ref{exp:W}) can be also obtained starting from the
characteristic function expressed in complex notation:
\begin{equation}
\hbox{Tr}\left[\varrho\left[(\ha_s^{\dag})^{h}\,\ha_t^{k}\right]_{\rm s}\right] =
  \left. (-1)^k \frac{\partial^{h+k}}{\partial \lambda_s^h\,\partial
    {\lambda_t^*}^k}\,\chi[\varrho](\boldsymbol{\lambda})
\right|_{\boldsymbol{\lambda}=\boldsymbol{0}}
\end{equation}
where $\chi[\varrho](\boldsymbol{\lambda}) = \hbox{Tr}[\varrho\,
D(\boldsymbol{\lambda})]$ and $D(\boldsymbol{\lambda})$ has been
defined in Eq.~(\ref{defDcmplx}): since its derivatives in the origin
of the complex plane generates symmetrically ordered moments of mode
operators, the characteristic function is also known as the
moment-generating function of the signal $\varrho$.
\par
In order to become more familiar with the covariance matrix formalism,
we consider the multi-mode state at thermal equilibrium at temperature
$T$ described by the density matrix $\nu = \bigotimes_{k=1}^n \:
\nu_{\rm th}(N_k)$ with:
\begin{align}
\nu_{\rm th}(N_k)
&=\frac{e^{-\beta_k \hat{a}^{\dag}_k \hat{a}_k}}
{\hbox{Tr}\left[e^{-\beta_k \hat{a}^{\dag}_k \hat{a}_k}\right]} =
\frac{N_k^{\hat{a}^{\dag}_k \hat{a}_k}}
{(1+N_k)^{\hat{a}^{\dag}_k \hat{a}_k+1}},\\
&= \frac{1}{1+{N}_k} \sum_{m=0}^\infty \left(\frac{{N}_k}{1
+{N}_k}\right)^m\: | m \rangle_k {}_k\langle m |
\label{th}\;,
\end{align}
where $\beta_k=\hbar\omega_k/(k_{\rm B}T)$, $k_{\rm B}$ being the
Boltzmann constant, and
$N_k=(e^{\beta_k}-1)^{-1}$ is the average number of quanta in the
$k$-th mode with frequency $\omega_k$. Its CM $\bmsigma_{\nu}$ turns
out to be diagonal and reads:
\begin{equation}\label{CM:th}
\bmsigma_{\nu} = \bigoplus_{k=1}^{n} \bmsigma_{\rm th}(N_k),
\end{equation}
where $\bmsigma_{\rm th}(N_k) = \frac12 (1+2N_k) \ii_{2}$ is the
$2\times 2$ CM of the $k$-th single-mode thermal state with $N_k$
average photons and $\ii_{m}$ is the $m\times m$ identity
matrix. Moreover, recalling that $\langle n |D(\lambda)| n \rangle =
e^{-\frac12 |\lambda|^2}L_n(|\lambda|^2)$, $L_n(z)$ being Laguerre
polynomials, one can easily calculate the expression of the
characteristic function of the thermal state (\ref{th}), that turns
out to be a Gaussian state. Note that if $N_k \to 0$ $\forall k$, then
$\bmsigma_{\nu} \to \frac12 \ii_{2n}$, that is the CM of the vacuum
state of $n$ bosons.
\par
More in general, the $2n\times 2n$ CM $\bmSigma_{\vset}$ of a
$n$-mode Gaussian state $\varrho_{\vset}$, $\vset = \{ 1,\ldots,n \}$,
can be re-written in the following block form:
\begin{equation}
\bmSigma_{\vset} =
\left(
\begin{array}{cccc}
\bmsigma_1 & \bmdelta_{12} & \cdots & \bmdelta_{1n} \\
\bmdelta_{12}^{\sT} & \bmsigma_2 & \cdots & \bmdelta_{2n} \\
\vdots & \vdots & \ddots & \vdots \\
\bmdelta_{1n}^{\sT} & \bmdelta_{2n}^{\sT} & \cdots & \bmsigma_n
\end{array}
\right),
\end{equation}
where $\bmsigma_k$ and $\bmdelta_{hk}$ are $2\times 2$ real
matrices. In particular, $\bmsigma_k$ corresponds to the CM of the
state $\varrho_k = \Tr_{\vset \setminus \{k\}}[\varrho_{\vset}]$ and
$\bmdelta_{hk}$ is related to the (classical or quantum) correlations
between the modes $h$ ad $k$: if $\bmdelta_{hk} = \bmzero$, then
$\varrho_{hk} = \Tr_{\vset \setminus \{h,k\}}[\varrho_{\vset}] =
\varrho_{h}\otimes \varrho_{k}$, that is the two modes are
uncorrelated, and the CM:
\begin{equation}\label{CM:hk}
\bmSigma_{hk} =
\left(
\begin{array}{cc}
\bmsigma_h & \bmdelta_{hk} \\
\bmdelta_{hk}^{\sT} & \bmsigma_k
\end{array}
\right)
\end{equation}
of the state $\varrho_{hk}$ reduces to the direct sum of the two
single-mode CMs, namely, $\bmSigma_{hk}= \bmsigma_h\oplus \bmsigma_k$.
%
%%%%%%%%%%%%%%%%%%%%%%%% SECTION %%%%%%%%%%%%%%%%%%%%%%%%%%
\section{Evolution of Gaussian states}
\label{s:symp:trasf}
When an evolution preserves the Gaussian character of a state, it can
be described with suitable transformations of the position- and
momentum-like operators or, equivalently, of $\hat{\bmR}$, that
preserve the commutation relations (\ref{defCom1}). These
transformations are called symplectic transformations and are the main
tool used to describe the kinematics of Gaussian states in the phase
space.
\par
First of all, we recall that the equations of motion of a classical
system of $n$ particles described by coordinates $\{q_1,\ldots,q_n\}$
and conjugated momenta $\{p_1,\ldots,p_n\}$ with Hamiltonian $H$ can be
summarized as:
\begin{eqnarray}
\dot R_k = \Omega_{kl}\, \frac{\partial H}{\partial R_l},
\label{HamEq1}\;
\end{eqnarray}
where $\boldsymbol{R} = (q_1,p_1,\ldots, q_n,p_n)^{\sT}$ and
$\dot{x}$ denotes time derivative and $\gr{\Omega}$ is the symplectic
matrix defined in \eq{defOM}. Given a transformation of coordinates
$\bmR \to \gr{R'} \equiv \gr{F}\bmR$, one has:
\begin{eqnarray}
\dot R'_k = F_{ks}\Omega_{st}F_{lt}\, \frac{\partial H}{\partial R_l'} 
\label{HamEq2}\;,
\end{eqnarray}
and thus the equations of motions remain invarians if and only if:
\begin{eqnarray}
\boldsymbol{F}\, \boldsymbol{\Omega} \boldsymbol{F}^{\sT} =
\boldsymbol{\Omega}\,,\quad \mbox{(symplectic condition)}
\label{SympCond}\;
\end{eqnarray}
which characterizes the symplectic transformations and, in turn,
describes the canonical transformations of coordinates.
\par
Form the quantum mechanical point of view, a mode transformation
$\hat{\gr{R'}}= \gr{F} \hat{\gr{R}}$ leaves the kinematics invariant
if it preserves canonical commutation relations (\ref{defCom1}): the
$2n\times 2n$ matrix $\gr{F}$ should satisfy the symplectic condition
(\ref{SympCond}). It is worth noting that if $\gr{F}$ and $\gr{G}$ are
symplectic transformations, then also $\gr{F}^{T}$,
$\gr{F}^{-1}=\bmOmega \gr{F}^T\bmOmega^{-1}$ and $\gr{F} \gr{G}$ are
symplectic: the set of the $2n \times 2n$ matrices satisfying the
condition Eq.~(\ref{SympCond}) form the symplectic group Sp$(2n,\rr)$.
\par
An important theorem due to J.~Williamson \cite{Wil36} guarantees that
every CM can be diagonalized through a symplectic transformation
\cite{dGro:06}. More in detail, if $\{d_k\}_{k=1}^{n}$ is the set of
the symplectic eigenvalues of the $2n \times 2n$ CM $\bmsigma$, namely
the moduli of the eigenvalues $\{\pm d_k\}_{k=1}^{n}$ of $i\bmOmega
\bmsigma$, where $\bmOmega$ is given in Eq.~(\ref{defOM}), then:
\begin{equation}\label{c3:SympDiagSG}
\bmsigma = \bmS\, \bmW \bmS^{\sT},
\end{equation}
where $\bmW=\bigoplus_{k=1}^n d_k\,\ii_2$ is a $n$-mode thermal state
with $N_k = d_k -\frac12$ average number of photon in the $k$-th mode
[see Eq.~(\ref{CM:th})], and $\bmS$ is the matrix which performs the
symplectic diagonalization (as we will see in the following, if
$\bmsigma$ describes a physical state, then $d_k\ge 1/2$, and, thus,
$N_k\ge 0$). Now, since the whole set of the symplectic
transformations is generated by Hamiltonians which are linear and
bilinear in the field modes \cite{SimonSympl:a,SimonSympl:b}, the
physical statement implied by decomposition (\ref{c3:SympDiagSG}) is
that every Gaussian state $\varrho$ can be obtained from a thermal
state $\nu$ by performing the unitary transformation $U_{\bmS}$
associated with the symplectic matrix $\bmS$, which, in turn, can be
generated by linear and bilinear interactions. Hence, the density
matrix corresponding to the decomposition (\ref{c3:SympDiagSG}) can be
written as:
\begin{equation}
\label{c3:GenericGS}
\varrho=U_{\bmS}\,\nu\, U_{\bmS}^\dag \;.
\end{equation}
\par
By using the uncertainty relation (\ref{HeisSG}), which is invariant
under the symplectic group Sp$(2n,\rr)$, and the decomposition
(\ref{c3:SympDiagSG}), we have:
\begin{equation}\label{c3:SympEigUncert}
\bmS\, \bmW \bmS^{\sT} + \frac{i}{2}\bmOmega \ge 0
\,\,\Rightarrow\,\,
\bmW \ge -\frac{i}{2}\bmOmega
\,\,\Rightarrow\,\,
d_k \ge\frac12,\, \forall k,
\end{equation}
that is the constraint on the CM by the uncertainty
relation leads to the constraints $d_k \ge 1/2$ on its symplectic
eigenvalues. From Eq.~(\ref{c3:purity:gauss:N}) it is
straightforward to see that a Gaussian state is pure if and only if
$d_k = 1/2$.
%
%%%%%%%%%%%%%%%%%%%%%%%% SECTION %%%%%%%%%%%%%%%%%%%%%%%%%%
\section{Linear and bilinear Hamiltonians}
\label{s:lin:bilin}
In order to preserve Gaussian states, a Hamiltonian should be linear
or bilinear in the fields mode \cite{schu:86}. This kind of
Hamiltonian can be experimentally realized by means of parametric
processes in quantum optical \cite{mandel,klysko}, optomechanical
\cite{pir:03,xia:10}, micromechanical \cite{cav08} and cold gases
\cite{meystre,PCB03,telebec1,telebec2,CPP04,khang} systems.  Though
the actual realization of these transformations necessarily involves
parametric interactions in nonlinear media, their quantum optical
implementation is often referred to as quantum information processing
with linear optics, according to the linearity of mode evolution.
\par
The most general Hamiltonian of this kind can be written as:
\begin{eqnarray}
H=\sum_{k=1}^{n} g_{k}^{(1)}\, \ha_k^\dag +
\sum_{k\ge l=1}^{n} g_{kl}^{(2)}\, \ha_k^\dag \ha_l+
\sum_{k,l=1}^{n} g_{kl}^{(3)}\, \ha_k^\dag \ha^\dag_l + h.c.,
\label{LinH}\;
\end{eqnarray}
and contains three main building blocks, which represent the
generators of the corresponding unitary evolutions to be
described in the following subsections. The mode transformation
imposed by the Hamiltonian (\ref{LinH}) and, thus, the evolution of the
vector $\hat{\bmR}$ and of the CM $\bmsigma$ writes:
\begin{equation}\label{modeLinH}
\hat{\gr{R}} \rightarrow \gr{F} \hat{\gr{R}}
+ \gr{d}\,,\quad \hbox{and} \quad
 \gr{\sigma} \rightarrow \gr{F}\, \bmsigma\, \gr{F}^{\sT},
\end{equation}
where $\gr{d}$ is a real vector and $\gr{F}$ a symplectic transformations.
Remarkably, the converse is also true, i.e., any symplectic
transformation of the form (\ref{modeLinH}) is generated by a unitary
transformation induced by Hamiltonians of the form (\ref{LinH})
\cite{SimonSympl:a,SimonSympl:b}. In this context, it is worth noting that a
useful decomposition of a generic symplectic transformation $\gr{F}$
is the following:
\be
\gr{F}=\gr{O}
\begin{pmatrix}
\gr{D} & \boldsymbol{0} \\
\boldsymbol{0} & \gr{D}^{-1}
\end{pmatrix}
\gr{O'}\;,\quad\mbox{(Euler decomposition) }
\label{c1:EulerDecomp}
\ee
where $\gr{O}$ and $\gr{O'}$ are orthogonal and symplectic
matrices, while $ \gr{D}$ is a positive diagonal matrix. The physical
implication of the Euler decomposition (\ref{c1:EulerDecomp}) is that
every symplectic transformation may be implemented by means of two
passive devices (described by the orthogonal matrices $\gr{O}$ and
$\gr{O'}$) and by single-mode squeezers (described by $\gr{D}$)
\cite{Bra99} to be addressed in the following subsections.
\subsection{Displacement operator and coherent states}
\label{ss:displacement}
The first block of the Hamiltonian in \eq{LinH} contains terms of the form
$H \propto g^{(1)}\:\ha^\dag + h. c. $ and is linear in the field modes. The
corresponding unitary transformations are the set of displacement
operators  we used in Sect.~\ref{s:notation} to define the
characteristic function. The comparison between
Eqs.~(\ref{displa_mode}) and (\ref{modeLinH}) shows that the CM is
left unchanged by the displacement operator while the first-moments
vector is displaced.
\par
The displacement operator is strictly connected with coherent states
\cite{gla:63}. For a single mode, coherent states are defined as the
eigenstates of the mode operator, i.e., $\ha|\alpha\rangle = \alpha
|\alpha\rangle$, where $\alpha \in \cc$ is a complex number. Using
Eq.~(\ref{displa_mode}), it can be shown that coherent states may be
defined also as $|\alpha\rangle = D(\alpha)|0\rangle$, that is the
unitary evolution of the vacuum through the displacement
operator. Properties of coherent states, such as the overcompleteness
and the nonorthogonality directly follow from those of the
displacement operator.
\par
Since the CM of coherent states is the same as the vacuum state one
[see Eq.~(\ref{CM:th}) with $N_k=0$], they are minimum uncertainty
states, i.e., they fulfill Ineq.~(\ref{HeisSG}) with equality sign
and, in addition, with uncertainties that are equal for the position-
and momentum-like operators (this is directly seen from the CM).
\subsection{Free evolution and two-mode mixing}\label{ss:BS}
The second block appearing in the Hamiltonian (\ref{LinH}), i.e.,
$\sum_{k\ge l=1}^{n} g_{kl}^{(2)}\, \ha_k^\dag \ha_l$, represents two
different physical processes.
\subsubsection{Phase shift}
The first process refers to the terms proportional to
$g^{(2)}\ha_k^{\dag}\ha_k$ and describes the free evolution of the
modes: in most cases these terms can be eliminated by choosing a
suitable interaction picture. The effect of free evolution is to add
an overall phase shift that, for single-mode fields, has no physical
meaning, but it is of extreme relevance in the case of interference
phenomena involving different beams of light, such as the
interferometric scheme used to implement the homodyne detection. The
evolution operator may be written as $U(\theta) =
\exp\{-i\theta\,\ha_k^{\dag}\ha_k\}$ and acts as a phase rotation on
the field mode $\ha_k$, i.e., $U^{\dag}(\theta)\,\ha_k\,U(\theta) =
e^{-i\theta}\,\ha_k$. Hence, the corresponding symplectic matrix
reads:
\begin{equation}\label{symp:PhR}
\boldsymbol{{\cal R}}_{\theta} = \left(
\begin{array}{cc}
\cos\theta & \quad \sin\theta \\ [1ex]
-\sin\theta & \quad \cos\theta
\end{array}\right)\,,
\end{equation}
and the evolution of the first-moments vector and the single-mode CM
follows form Eqs.~(\ref{modeLinH}), with $\gr{d}=\gr{0}$.
\subsubsection{Two-mode mixing}
The second process, involving different mode operators, describes a
linear mixing of two modes and, in the quantum optics context, the
simplest example corresponds to a Hamiltonian of the form $H\propto
\hat{a}^\dag \hat{b} + \hat{b}^\dag \hat{a}$, where for the sake of
simplicity we consider a system of two modes $\hat{a} \equiv\hat{a}_1$
and $\hat{b} \equiv\hat{a}_2$. This Hamiltonian describes the
action of a beam splitter, i.e., the interaction taking place in a
linear optical medium such as a dielectric plate.  The evolution
operator can be recast in the form:
\begin{eqnarray}
  U(\zeta) = \exp\left\{\zeta \hat{a}^\dag \hat{b} -
\zeta^* \hat{a} \hat{b}^\dag\right\}\:,
\label{ubs}\;
\end{eqnarray}
where the coupling $\zeta = \phi\, e^{i\theta} \in {\mathbb C}$ is
proportional to the interaction length (time) and to the linear
susceptibility of the medium. The two-mode mixer is a ``passive''
device, i.e., the total number of quanta in the two modes is a
constant of motion.
\par
The Heisenberg evolutions of modes $\ha$ and $\hb$ are given by:
\begin{subequations}\label{EvolMode}
\begin{align}
U^\dag(\zeta)\: \hat{a}\: U(\zeta)&=
 \cos\phi \:\hat{a} + e^{i\theta}\sin\phi\:\hat{b} \,, \\
U^\dag(\zeta)\: \hat{b}\: U(\zeta)&=
 \cos\phi\:\hat{b} -  e^{-i\theta}\sin\phi\:\hat{a}\,,
\end{align}
\end{subequations}
respectively, and the corresponding symplectic matrix $\gr{S}_\zeta$
reads \cite{FOP:05}:
\begin{equation}\label{symp:BS}
\gr{S}_{\zeta} = \left(\begin{array}{cc}
\cos\phi\, \ii_2 & \quad \sin\phi\, \boldsymbol{{\cal R}}_{\theta}\ \\[1ex]
-\sin\phi\, \boldsymbol{{\cal R}}_{\theta}^{\sT} & \quad \cos\phi\, \ii_2
\end{array}
\right)\,,
\end{equation}
where the $2\times 2$ matrix $\boldsymbol{{\cal R}}_{\theta}$ is
defined in Eq.~(\ref{symp:PhR}).  The first-moments vector and
two-mode CMs evolve as usual according to Eqs.~(\ref{modeLinH}), only
with $\gr{d}=\gr{0}$.

\subsection{Single-mode squeezing}
\label{ss:sm:squeezing}
In the particular case of quantum optics, the last block of the
Hamiltonian (\ref{LinH}) describes $\chi^{(2)}$ nonlinear interactions
in which a photon in the input (pump) is converted into two photons,
conserving both the energy and the momentum. If the so-called
phase-matching conditions are arranged in order to emit the two
photons into the same mode $\ha$, we obtain the single-mode squeezing
transformations, which, thus, correspond to Hamiltonians of the form
$H\propto (\ha^{\dag})^{2} + h.c. $ \cite{marc:08}. Squeezing has been
firstly introduced for quadrature operators and refers to a phenomenon
in which an observable or a set of observables exhibit a second moment
below the corresponding vacuum level \cite{yuen76}.
\par
The single-mode squeezing operator is usually written as:
\begin{eqnarray}
S (\xi) =
\exp\left\{\mbfrac \left[\xi (\ha^{\dag})^{2} - \xi^* \ha^2\right]\right\}\:,
\label{usq}
\end{eqnarray}
which corresponds to the following mode evolutions:
\begin{subequations}\label{modesq}
\begin{align}
  S^\dag(\xi) \: \ha \: S(\xi) &=
\cosh r\: \ha + e^{i\psi}\sinh r\: \ha^\dag\,, \\
S^\dag(\xi) \: \ha^\dag \: S(\xi) &=
\cosh r\: \ha^\dag + e^{-i\psi}\sinh r \:\ha \;,
\end{align}
\end{subequations}
with $\xi=r e^{i\psi}$.  By using the mode transformation in
Eqs.~(\ref{modesq}) and the definition of the quadrature operators
(\ref{defQP}), it is straightforward to find the symplectic
transformation $\gr{\Sigma}_\xi$ describing the single-mode squeezing,
namely:
\begin{eqnarray}
\gr{\Sigma}_\xi =  \cosh r\, \ii_2 + \gr{R}_\xi\, \quad 
\mbox{with} \quad
\gr{R}_\xi = \sinh r
\left(\begin{array}{cc} \cos\psi & \quad \sin\psi \\ [1ex]
\sin\psi & \quad-\cos\psi \end{array}\right)
\label{Smatrix}\;,
\end{eqnarray}
which allows to calculate the evolution of the first-moments vector
and CM according to Eqs.~(\ref{modeLinH}), but with $\gr{d}=\gr{0}$.

\subsection{Two-mode squeezing}\label{ss:tm:squeezing}
The two-mode squeezing transformations correspond to Hamiltonians of
the form $H\propto \ha^\dag \hb^\dag + h.c. $ and describe
$\chi^{(2)}$ nonlinear interactions introduced in the previous
subsection but with the two photons emitted in different modes. The
evolution operator is usually written as:
\begin{eqnarray}
S_2 (\xi) = \exp\left\{\xi \ha^{\dag} \hb^{\dag} - \xi^* \ha \hb\right\}\:.
\label{usq2}
\end{eqnarray}
The corresponding evolutions of the two modes read:
\begin{subequations}\label{modesq2}
\begin{eqnarray}
S_2^\dag(\xi) \: \ha \: S_2(\xi) &=
\cosh r \: \ha + e^{i\psi} \sinh r \: \hb^{\dag}\,, \\
S_2^\dag(\xi) \: \hb \: S_2(\xi) &=
\cosh r \: \hb - e^{i\psi} \sinh r \: \ha^{\dag}\,,
\end{eqnarray}
\end{subequations}
where $\xi=re^{i\psi}$ and the symplectic transformation associated
with the two-mode squeezer is represented by the block matrix:
\begin{eqnarray}
\gr{\Sigma}_{2\xi} =\left(
\begin{array}{cc} 
 \cosh r\, \ii_2 & \gr{R}_\xi\\ [1ex]
 \gr{R}_\xi & \cosh r\, \ii_2 
\end{array}
\right)\,,
\label{Qxi}
\end{eqnarray}
where $\gr{R}_\xi$ is defined as in (\ref{Smatrix}) and, as usual, the
CM $\gr{\sigma}$ and the first-moments vector of a bipartite input
state transform according to Eqs.~(\ref{modeLinH}) with
$\gr{d}=\gr{0}$.

%%%%%%%%%%%%%%%%%%%%%%%% SECTION %%%%%%%%%%%%%%%%%%%%%%%%%%
\section{Single-mode Gaussian states}
\label{sec:sm:GS}
\begin{figure}
\resizebox{0.32\columnwidth}{!}{
\includegraphics{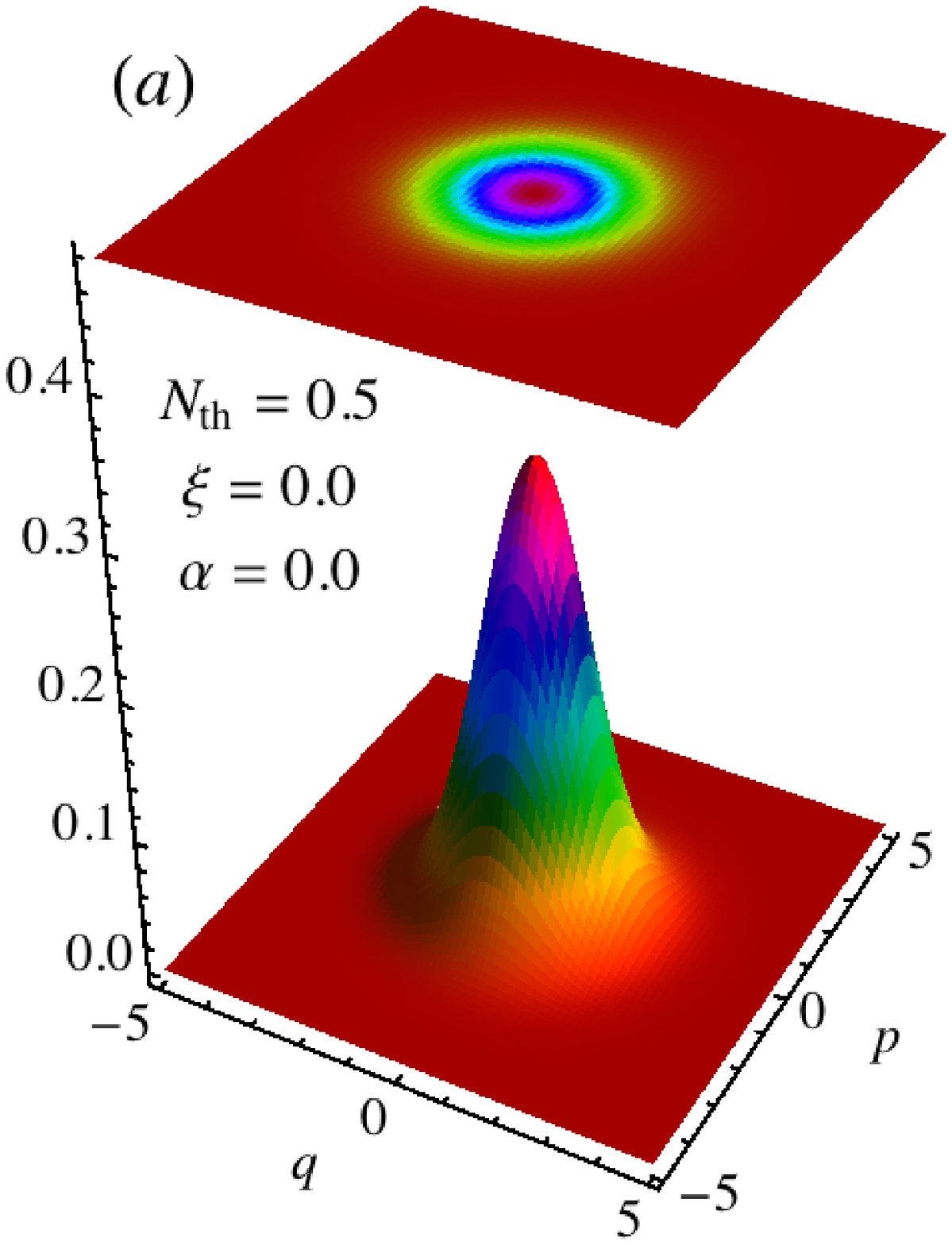} }
\resizebox{0.32\columnwidth}{!}{
\includegraphics{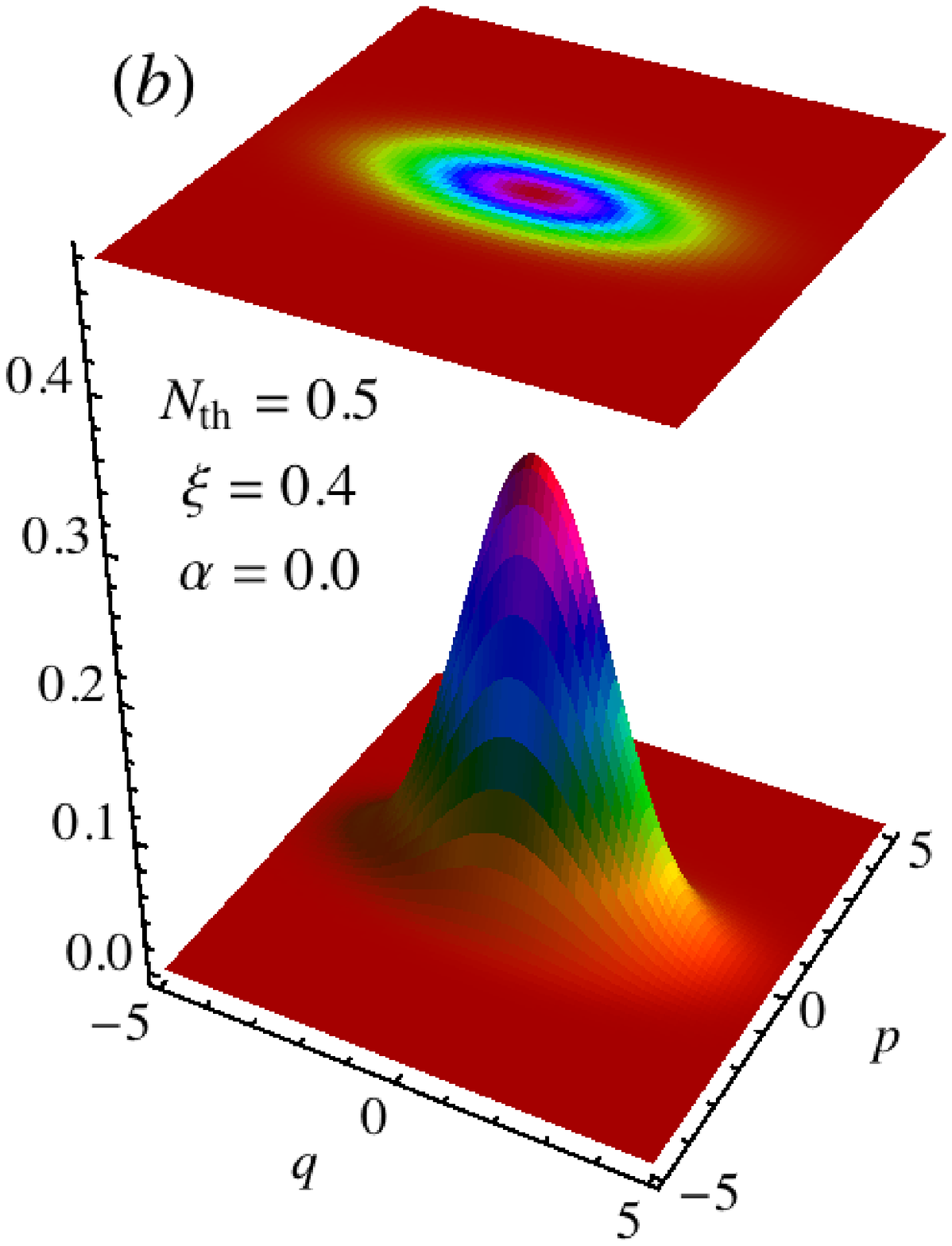} }
\resizebox{0.32\columnwidth}{!}{
\includegraphics{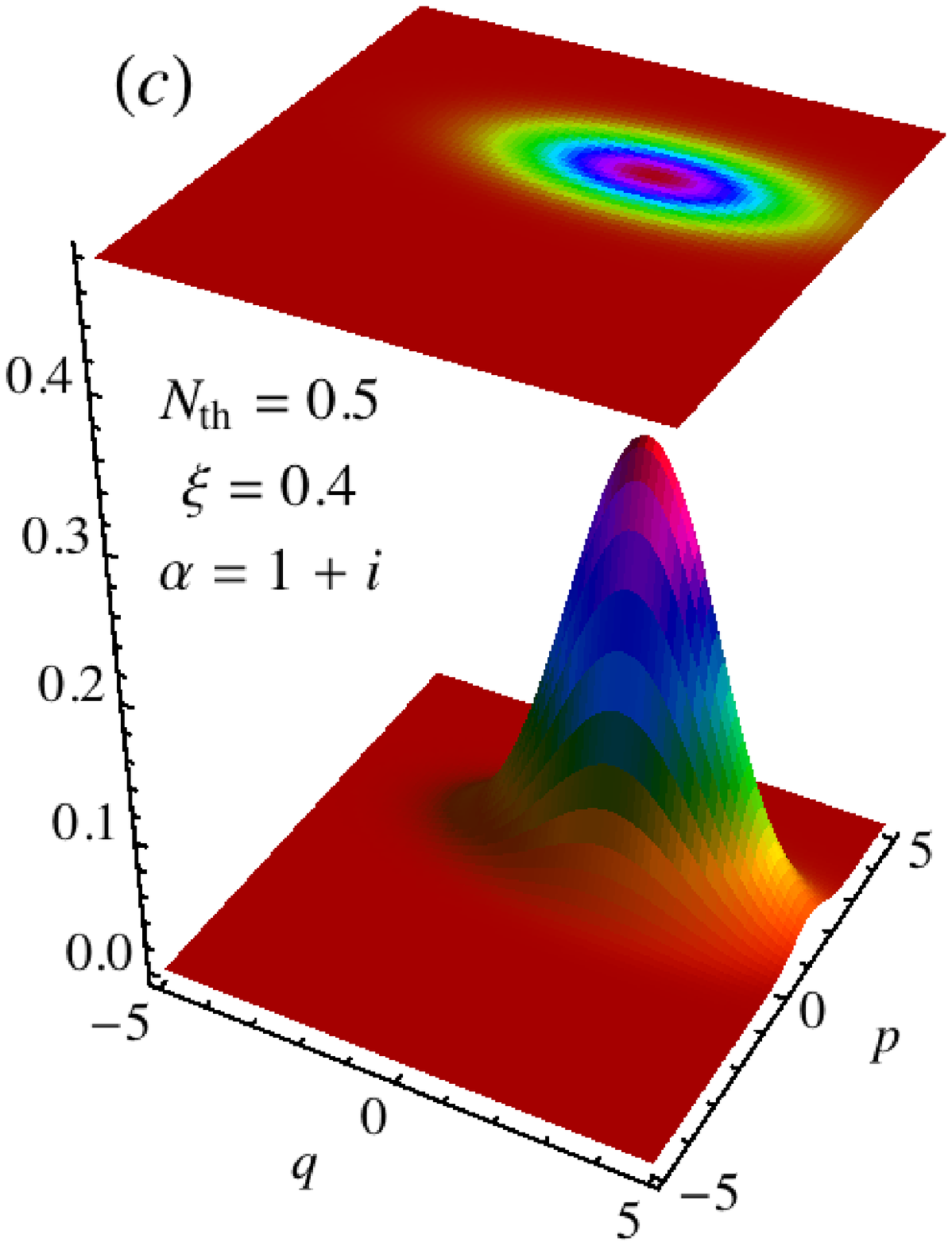} }
\caption{\label{f:W:GS} (Color online) Plots of the Wigner function of
  the single-mode Gaussian state $\varrho = D(\alpha)\,S(\xi)\,
  \nu(N_{\rm th})\, S^{\dag}(\xi)\, D^{\dag}(\alpha)$, for different
  values of $N_{\rm th}$, $\xi$ and $\alpha$: ($a$) thermal state,
  ($b$) squeezed thermal state and (c) displaced squeezed
  thermal state.}
\end{figure}
In the case of a single-mode Gaussian state, Eq.~(\ref{c3:GenericGS})
may be written as \cite{Ada95}:
\begin{eqnarray}
\varrho= D(\alpha)S(\xi)\,\nu_{\rm th}(N)\,S^{\dag}(\xi)
D^{\dag}(\alpha) \,,
\label{c3:rho:1m}\;
\end{eqnarray}
and the corresponding CM and first-moments vector can be easily
calculated by applying the phase-space analysis introduced in
Sect.~\ref{s:lin:bilin}. In particular the CM can be calculated
applying the squeezing transformation (\ref{Smatrix}) to the CM
$\bmsigma_{\rm th}(N)$ of the thermal state $\nu_{\rm th}(N)$, namely
$\bmsigma = \bmSigma_\xi\:\bmsigma_{\rm th}(N)\:
\bmSigma_\xi^{\sT}$. The explicit expressions of its elements are:
\begin{subequations}
\label{c3:vars}
\begin{align}
\sigma_{kk}&= \frac{1+2N}{2}\:
\Big[\!\cosh(2r)-(-1)^{k}\sinh(2r)\cos\psi\Big] \:,
\quad (k=1,2) \\
\sigma_{12}&=\sigma_{21}=\frac{1 + 2N}{2}\:
\sinh(2r)\sin\psi \:,
\end{align}
\end{subequations}
with $\xi = r e^{i\psi}$, while the first-moments vector reads
$\langle\hat{\bmR}\rangle=
\sqrt{2}(\re{\alpha},\im{\alpha})^{\sT}$. By using
Eq.~(\ref{c3:purity:gauss:N}) we can calculate the purity of the
Gaussian state (\ref{c3:rho:1m}), that reads $\mu(\varrho)=(1+2
N)^{-1}$: the purity of a generic single-mode Gaussian state depends
only on the average number of thermal photons, as one should expect
since displacement and squeezing are unitary operations and, thus, do
not affect the purity of a state. The same observation holds when we
address the von Neumann entropy:
\begin{equation}\label{vN:ent}
S_V(\varrho) = -\hbox{Tr}[\varrho \ln
\varrho]\,.
\end{equation}
For a single-mode Gaussian state we have:
\begin{equation}
\label{c3:S_V1m}
S_V(\varrho) = f\left(\sqrt{\det[\bmsigma]}\right)\,,
\end{equation}
where:
\begin{equation}\label{f:x}
f(x)=
\left(x+\mbox{$\frac12$}\right)\ln\left(x+\mbox{$\frac12$}\right)
-\left(x-\mbox{$\frac12$}\right)\ln\left(x-\mbox{$\frac12$}\right)\,,
\end{equation}
and $\sqrt{\det[\bmsigma]} = N+\frac12 = [2\mu(\varrho)]^{-1}$ corresponds to
the only symplectic eigenvalue of the $2\times 2$ positive-definite
symmetric matrix $\bmsigma$, since, in this case, the two eigenvalues
of $i\bmOmega \bmsigma$ are $d_{\pm} = \pm\sqrt{\det[\bmsigma]}$, as
follows from the Williamson's theorem applied to a single-mode CM.
\par
Starting from Eqs.~(\ref{c3:rho:1m}) and (\ref{c3:vars}) one can
obtain the CMs of the coherent state (by setting $N_{\rm th}=\xi=0$)
or of the squeezed vacuum state (with $\alpha = \xi=0$). In
Fig.~\ref{f:W:GS} the Wigner function of the Gaussian state
(\ref{c3:rho:1m}) is plotted for different values of the involved
parameters.

%%%%%%%%%%%%%%%%%%%%%%%% SECTION %%%%%%%%%%%%%%%%%%%%%%%%%%
\section{Two-mode Gaussian states}
\label{sec:tm:GS}
We can identify different classes of two-mode or, more in general,
bipartite Gaussian states. However, since these states are the
simplest scenario where to investigate the fundamental issue of
entanglement in quantum information, it is useful to introduce
equivalence classes of Gaussian states with the same amount of
entanglement, i.e., Gaussian states whose CMs are connected by local
symplectic transformations and, thus, are locally equivalent.
If we write the CM of a two-mode Gaussian state as:
\begin{equation}
\label{c3:CM2pGeneric}
\bmsigma = \left(
\begin{array}{cc}
\bmA   & \bmC \\
\bmC^{\sT} & \bmB
\end{array}
\right) \,,
\end{equation}
where $\bmA$, $\bmB$ and $\bmC$ are $2\times 2$ matrices, then we can
define four local symplectic invariants, i.e., quantities
that are left unchanged by local symplectic transformations:
\begin{equation}\label{loc:symp:inv}
I_1=\det[\bmA],\quad I_2=\det[\bmB],\quad
I_3=\det[\bmC],\quad  I_4=\det[\bmsigma]\,.
\end{equation}
The CMs of locally equivalent states can be reduced to the following
standard or normal form \cite{BR03,GEC+03}:
\begin{equation}
\label{c3:NF2mMix}
\bmsigma = 
\left(
\begin{array}{cccc}
 a   &  0   &  c_1 &  0   \\
 0   &  a   &  0   &  c_2 \\
 c_1 &  0   &  b   &  0   \\
 0   &  c_2 &  0   &  b 
\end{array}
\right) \,,
\end{equation}
where the values of $a$, $b$, $c_1$, and $c_2$ are determined by the
local symplectic invariants (\ref{loc:symp:inv}), namely $a^2=I_1$,
$b^2=I_2$, $c_1c_2=I_3$ and $(ab-c_1^2)(ab-c_2^2)=I_4$.
\par
The two symplectic eigenvalues of the CM of a generic two-mode
Gaussian state can be computed in terms of the symplectic invariants
\cite{SIS03}:
\begin{equation}
\label{c3:SympEig2m}
d_{\pm} = \sqrt{
  \frac{\Delta(\bmsigma)\pm\sqrt{\Delta(\bmsigma)^2-4I_4}}{2} }\,,
\end{equation}
with $\Delta(\bmsigma) = I_1+I_2+2I_3$ and, in turn, the uncertainty
relation (\ref{c3:SympEigUncert}) reduces to:
\begin{equation}
\label{c3:HeisSympEig} 
d_-\ge 1/2 \;.
\end{equation}
Note that for a pure two-mode Gaussian state we have $I_4=1/16$ and
$\Delta(\bmsigma)=1/2$, i.e., a pure Gaussian state has minimum
uncertainty. Moreover, bipartite pure states necessarily have a
symmetric normal form, i.e., $a=b$ in Eq.~(\ref{c3:NF2mMix}), as can
be seen by equating the entropies of the subsystems.
\par
A relevant subclass of Gaussian states is formed by the two-mode
squeezed thermal states (for a general parameterization of an
arbitrary bipartite Gaussian state, by means of a proper symplectic
diagonalization, see Ref.~\cite{SIS03}), i.e., states generated by
applying the two-mode squeezing operator to a two-mode thermal state,
namely: \be
\label{c3:thermalsq}
\varrho =
S_{2}(\xi)\,
\nu_{\rm th}(N_1)\otimes\nu_{\rm th}(N_2)\,
S_{2}^{\dag}(\xi)\,.
\ee
We can calculate the CM of the state (\ref{c3:thermalsq}) as
$\bmsigma=\gr{\Sigma}_{2\xi} \bmsigma_\nu \gr{\Sigma}_{2\xi}^{\sT}$,
where $\gr{\Sigma}_{2\xi}$ is the symplectic two-mode squeezing matrix
(\ref{Qxi}) and $\bmsigma_\nu$ is the CM of the thermal state $\nu$
given in \refeq{CM:th} with $n=2$. In formula,
\begin{equation}
\bmsigma = \frac{1}{2}\begin{pmatrix}
A\, \mathbbm{1}_2&  C\, \boldsymbol{R}_\xi \\
C\, \boldsymbol{R}_\xi & B\, \mathbbm{1}_2 
\end{pmatrix}\,,
\label{c:matrix}
\end{equation}
$\boldsymbol{R}_\xi$ being defined in \refeq{Qxi} and, if we assume
$\xi = r \in \rr$:
\begin{subequations}\label{c3:thermalsqCov}
\begin{align}
A &\equiv A(r, N_1, N_2) =
(1+N_1+N_2)\cosh (2r) + (N_1-N_2)\,,\\
B &\equiv B(r, N_1, N_2) =
(1+N_1+N_2)\cosh (2r) - (N_1-N_2)\,,\\
C &\equiv C(r, N_1, N_2) =
\left( 1 + {N_1} + {N_2} \right)\,\sinh (2r) \,.
\end{align}
\end{subequations}
In particular, if $N_1 = N_2 = 0$ we have so-called twin-beam state
(TWB) or two-mode squeezed vacuum, that plays a leading role in
quantum information with continuous variable. The first name, TWB,
refers to the fact that it shows perfect correlation in the photon
number, i.e., it is an eigenstate of the photon number difference
$\ha^\dag \ha - \hb^\dag \hb$, which is a constant of motion as the reader can
verify. The second name is instead connected with a duality under the
action of a balanced beam splitter, since one has:
\begin{eqnarray}
U^\dag(\mbox{$\frac\pi4$}\, e^{i\theta}) \:S_2(\xi)\:
U(\mbox{$\frac\pi4$}\,e^{i\theta})
= S(\xi e^{i\theta}) \otimes 
S(-\xi e^{-i\theta})
\label{duality}\;,
\end{eqnarray}
where $U(\zeta=\frac{\pi}{4}\, e^{i\theta})$ is the evolution operator
of Eq.~(\ref{ubs}) for a balanced beam splitter, $S_2(\xi)$ is
the two-mode squeezing operator (\ref{usq2}), and $S(\xi)$
is the single-mode squeezing operator of Eq.~(\ref{usq}) acting on the
evolved mode out of the mixer. In other words, a TWB entering a
balanced beam splitter evolves into a factorized state composed of two
squeezed vacua with opposite squeezing phases \cite{joint} and,
{\em viceversa}, a TWB may be generated by using single-mode squeezers and a
linear mixer as in the first continuous variable teleportation
experiment \cite{furu}.

\subsection{Entropies  and mutual information}
\subsubsection{Von Neumann entropy}
As we have seen in Sect.~\ref{s:symp:trasf}, a consequence of
Williamson's theorem is that every Guassian state can be generated
acting on a thermal state with unitary transformations. Thus, the von
Neumann entropy of a generic Gaussian state reduces to that of the
thermal state obtained from it by symplectic diagonalization, since
unitary operations do not affect the entropy of the whole state. In
the case of a two-mode Gaussian state $\varrho_{AB}$ with CM
$\bmsigma$, using Eq.~(\ref{c3:S_V1m}) and the additivity of von
Neumann entropy (\ref{vN:ent}) for tensor product states, i.e.,
$S_V(\varrho_A\otimes\varrho_B) = S_V(\varrho_A) + S_V(\varrho_B)$, we
obtain:
\begin{equation}
\label{c3:S_V2m}
S_{V}(\varrho_{AB}) = f(d_+)+f(d_-) \;,
\end{equation}
where $f(x)$ has been defined in Eq.~(\ref{f:x}) and $d_{\pm}$ are the
symplectic eigenvalues of $\bmsigma$ written explicitly in
\refeq{c3:SympEig2m}.
\subsubsection{Mutual information and conditional entropies}
Starting from the von Neumann entropies of the state $\varrho_{AB}$
and of the two subsystems $\varrho_A=\Tr_B[\varrho_{AB}]$ and
$\varrho_B=\Tr_A[\varrho_{AB}]$ it is possible to assess how much
information about $\varrho_{AB}$ one can obtain by addressing the
single parties. This is of course related to the correlations between
the two modes and can be quantified by means of the quantum mutual
information or the conditional entropies \cite{slep:71}.
\par
The quantum mutual information is defined as:
\begin{equation}\label{mut:info}
I_M\left( \varrho_{AB} \right) = S_{V}\left( \varrho_{A}\right) +
S_{V}\left( \varrho_{B}\right) - S_{V}\left( \varrho_{AB} \right),
\end{equation}
and can be easily expressed in terms of the symplectic invariants
(\ref{loc:symp:inv}) of ${\boldsymbol{\sigma}}$ and its symplectic
eigenvalues (\ref{c3:SympEig2m}) as follows:
\begin{equation}
I_M(\varrho_{AB})=f\left( \sqrt{I_1}\right) +f\left( \sqrt{I_2}\right)
-f(d_{+})-f(d_{-}).  \label{mutual info}
\end{equation}
Note that $f\left( \sqrt{I_{1}}\right) = S_V(\varrho_{A})$ and
$f\left( \sqrt{I_{2}}\right) = S_V(\varrho_{B})$, since $\varrho_{A}$
and $\varrho_{b}$ are a single-mode Gaussian states (see
Sect.~\ref{sec:sm:GS}). It is also worth noting that, in the case of
pure states, the entropies $S_{V}\left( \varrho_{A}\right) =
S_{V}\left( \varrho _{B}\right)$ correspond to the unique measure of
entanglement for pure bipartite states \cite{pop:97}.
\par
The conditional entropies are defined as: 
\begin{subequations}\label{cond:ent}
\begin{align}
S_{A|B}(\varrho_{AB})& =S_V(\varrho_{AB} )-S_V(\varrho_{B})\,,
\nonumber\\
&= f(d_{+}) + f(d_{-}) - f\left(\sqrt{I_2}\right)\,,\label{cond12} \\
S_{B|A}(\varrho_{AB})& =S_V(\varrho_{AB} )-S_V(\varrho_{A})
\nonumber\\
&= f(d_{+}) + f(d_{-}) - f\left(\sqrt{I_1}\right)\,,\label{cond21}
\end{align}
\end{subequations}
and can also assume negative values.  If $S_{A|B}(\varrho_{AB})\geq 0$, the
conditional entropy gives the amount of information that the party $A$
should send to the party $B$ in order to allow for the full knowledge
of the overall state $\varrho_{AB}$. If $S_{A|B}(\varrho_{AB})<0$, the
party $A$ does not need to send any information to the other and, in
addition, they gain $-S_{A|B}(\varrho_{AB})$ bits of entanglement,
respectively [analogous considerations hold for
$S_{B|A}(\varrho_{AB})$]. This has been proved for the case of
discrete variable quantum systems \cite{horo:05} and conjectured
\cite{gen:08} for infinite dimensional ones.

\subsection{Separability of Gaussian states}
\label{ss:separability}
A bipartite state $\varrho_{AB}\in {\cal H}_{A}\otimes{\cal H}_{B}$ is
separable if it can be written as a convex combination of product
states \cite{wer89}, namely, $ \varrho_{AB}=\sum_k p_k \varrho_k^{(A)}
\otimes \varrho_k^{(B)}$ where $p_k\ge0$, $\sum_k p_k=1$, and
$\varrho_k^{(h)}\in {\cal H}_h$, $h=A,B$. Finding the convex
combination of a separable state is a challenging task; nevertheless
the separability can be revealed with the aid of positive but not
completely positive maps. In particular, positivity under partial
transposition ({\tt ppt}), that is the transposition applied only to a
part of a system, has been introduced in entanglement theory by
A.~Peres \cite{Per96} as a necessary condition for separability. In
fact, if we apply, for instance, transposition only to elements of the
first subsystem $A$ of a separable state $\varrho_{AB}$, we have
$\varrho_{AB}^{\scriptscriptstyle T_A}=\sum_k p_k
\big(\varrho_k^{\scriptstyle (A)}\big)^{\sT} \otimes
\varrho_k^{\scriptstyle (B)}$. Now, since
$\big[\varrho_k^{(A)}\big]^{\sT}=\big[\varrho_k^{(A)}\big]^*$,
transposition corresponds to complex conjugation and the transposed
matrix is a legitimate density matrix itself, being non-negative,
self-adjoint and with unit trace. Then none of the eigenvalues of
$\varrho^{\scriptscriptstyle T_A}$ is negative if $\varrho$ is
separable. The {\tt ppt} criterion is usually only necessary and
entangled states with positive partial transposed density matrix are
known to exist and are called bound-entangled states
\cite{hor:97}. R.~Simon, however, has proved that for two-mode
Gaussian states it represents also a sufficient condition for
separability \cite{Sim00}.
\par
Since complex conjugation corresponds to time reversal of the
Schr\"{o}dinger equation, in terms of continuous variables
transposition corresponds to a sign change of the momentum variables,
i.e., a mirror reflection.  For a two-mode system described by the
density matrix $\varrho_{AB}$, partial transposition with respect to
system $A$ will be performed on the phase space through the action of
the matrix $\bmDelta_A= {\rm Diag}(1,-1) \oplus {\mathbbm 1}_2$, where
the first factor of the direct sum, representing the mirror
reflection, refers to subsystem $A$ and the second one to subsystem
$B$ (partial transposition with respect to subsystem $B$ is obtained
in a similar way). Hence, the positivity of the partial
transposed operator leads to the following uncertainty relation:
\begin{eqnarray}
\widetilde{\bmsigma} +
\frac{i}{2}\, \boldsymbol{\Omega} \geq 0\,, \quad \hbox{or}\quad
 \bmsigma \ge
-\frac{i}{2}\, {\widetilde {\boldsymbol \Omega}}_A\,,
\label{c4:HeisSG}
\end{eqnarray}
where $\widetilde{\bmsigma} = \bmDelta_A\: \bmsigma\:\bmDelta_A$ and
${\widetilde {\boldsymbol
    \Omega}}_A=\bmDelta_A\:\boldsymbol{\Omega}\:\bmDelta_A$.
Furthermore, recalling the definition (\ref{loc:symp:inv}) of the four
local symplectic invariants, now we have:
\begin {equation}
{\tilde I}_1=I_1\,,
\qquad {\tilde I}_2=I_2\,, \qquad {\tilde I}_3=-I_3\,, \qquad {\tilde
  I}_4=I_4\,,
\end{equation}
where ${\tilde I}_k$ are the symplectic invariants referred to
$\widetilde{\bmsigma}$. Thus, in terms of the symplectic eigenvalues
${\tilde d}_{\pm}$ of the partially transposed CM the {\tt ppt}
criterion reduces to: \be
\label{c4:nptSympEig} {\tilde d}_{-}\ge 1/2 \;.
\ee
with:
\be
\label{c3:SympEig2m:ppt}
\tilde{d}_{\pm} = \sqrt{
  \frac{\widetilde{\Delta}(\bmsigma)\pm
\sqrt{\widetilde{\Delta}(\bmsigma)^2-4I_4}}{2} }\;,
\ee
where $\widetilde{\Delta}(\bmsigma)=I_1+I_2-2I_3$.
Here we have shown that the {\tt ppt} criterion is necessary for
separability. As for its sufficiency we refer to the original
paper \cite{Sim00}.
\par
An equivalent necessary and sufficient criterion is based on the
evaluation of the sum of the variances associated with a pair of
EPR-like operators, defined on the two different subsystems
\cite{dua00}. The insight underlying this criterion is that for an
entangled state it is possible to gain information on one of the
subsystems suitably measuring the other one. This criterion leads to
an inequality that can be expressed in terms of elements of the CM
expressed in the standard form (\ref{c3:NF2mMix}), namely:
\begin{equation}
\tilde{a} \gamma^{2}+\frac{\tilde{b}}{\gamma^{2}}-
\left\vert \tilde{c}_{1}\right\vert
-\left\vert \tilde{c}_{2}\right\vert
-\left( \gamma^{2}+\frac{1}{\gamma^{2}} \right) < 0,
\end{equation}%
where we introduced the quantities:
$\gamma^{2}=\sqrt{(\tilde{b}-1/2)/(\tilde{a}-1/2)}$, $\tilde{a} = 2 a
\cosh 2r_1$, $\tilde{b} = 2 b \cosh 2r_2$, $\tilde{c}_1 = 2 c_1
\exp{(r_1+r_2)}$, $\tilde{c}_2 = 2 c_2 \exp{[-(r_1+r_2)]}$, and $r_1$
and $r_2$ are suitable squeezing parameters to transform the CM
(\ref{c3:NF2mMix}) into the so-called Duan canonical form (see
Ref.~\cite{dua00} for details). A separable state, whether Gaussian or
not, will not satisfy the above inequality.
\par
It is worth noting that also the negativity of the conditional
entropies (\ref{cond:ent}) is a sufficient condition for entanglement
\cite{cerf:99}.

\subsection{On the quantification of Gaussian entanglement}
\label{ss:quant:ent}
For a two-mode state, a quantitative measure of entanglement can be given 
on the observation that the larger is the violation $\tilde{d}_{-}<1/2$ 
the stronger is the entanglement, or more properly, the stronger the 
resilience of entanglement to noise \cite{sal1,sal2,nm1,nm2}. 
The logarithmic negativity for a two-mode Gaussian state,
is given by \cite{vid02}:
\begin{equation}
E(\bmsigma)=\max \left\{ 0,-\log 2\tilde{d}_{-} \right\} ,
\end{equation}%
and it is a simple increasing monotone function of the minimum
symplectic eigenvalue $\tilde{d}_{-}$ (for $0<\tilde{d}_{-}<1/2$).
Thus, it represents a good candidate for evaluating entanglement in a
quantitative way.
\par
Another convenient and useful way of looking at the entanglement
evolution in continuous variable systems is by means of the
entanglement of formation (EoF), which corresponds to the minimal
amount of entanglement of any ensemble of pure bipartite states
realizing the given state \cite{EoF,G:03}. In general the derivation
of an expression of the EoF for arbitrary states is not a simple task.
\par
In the case of a symmetric bipartite Gaussian state with CM
given by \eqref{c3:NF2mMix} with $a=b$, the EoF reads \cite{G:03}:
\begin{equation}\label{EoF}
E_{F} =f(x_m),
\end{equation}
where $f(x)$ is defined in Eq.~(\ref{f:x}), $x_m = (\tilde{d}_{-}^{2}
+ 1/4)/(2 \tilde{d}_{-})$, and $\tilde{d}_{-}$ is the minimum
symplectic eigenvalue of the partially transposed CM given in
Eq.~\eqref{c3:SympEig2m}.
\par
For the two-mode squeezed thermal state (\ref{c3:thermalsq}), in which
the standard form of the CM is obtained from Eq.~(\ref{c3:NF2mMix})
with $a\ge b$ and $c_1 = -c_2 = c\ge 0$, the EoF is still given by
Eq.~(\ref{EoF}) but with \cite{EoF2008}:
\begin{equation}
x_m = \frac{(a+b)(ab - c^2 + \hbox{$\frac14$})-
2c\sqrt{\det(\bmsigma+\frac{i}{2}\bmOmega)}}
{(a+b)^2-4c^2}.
\end{equation}
The EoF of other classes of two-mode Gaussian states can be evaluated by
following the general prescription proposed in Ref.~\cite{EoF2008}.
\par
Quantitative estimation of entanglement can be also obtained by
means of entropy functionals \cite{marc:08}. In particular, the
degree of entanglement of an ideal bipartite system can be assessed
following the analysis presented in \cite{marc:93}.
%%%
\subsection{Gaussian quantum discord}
\label{ss:quant:disc}
The correlations of a bipartite quantum system $\varrho_{AB}$,
quantified by the mutual information (\ref{mut:info}), can be divided
in a quantum part, known as quantum discord, and a classical part
\cite{zur:01}. The classical correlations are defined as the maximum
amount of information we can gain on one part of the system by locally
measuring the other subsystem, and, thus, can be written as a function
of the von Neumann entropies of the two subsystems as follows
\cite{HenVed}:
\begin{equation}
    \mathcal{C}_{A|B}(\varrho_{AB})=
\max_{\Pi_k}\bigl\{S_{V}(\varrho_{A})-\sum_k p_k
    S_{V}(\varrho^{\Pi_k}_{A|B})\bigl\},
\label{ClassCorrs}\end{equation}
where the set $\{ \Pi_k \}$, $\Pi_k\ge 0$ and $\sum_k \Pi_k
=\mathbbm{I}$, represents a positive operator-valued measure
(POVM), $\varrho^{\Pi_k}_{A|B}=\mbox{Tr}_B[\varrho_{AB}\:
\mathbbm{I}\otimes\Pi_k]/p_{k}$ is the conditional state of subsystem
$A$ when the $k$-th outcome occurs in a measurement of subsystem $B$
and $p_k=\mbox{Tr}_{AB}[\varrho_{AB}\: \mathbbm{I}\otimes\Pi_k]$. The
maximum is taken over all the POVMs performable on one
subsystem. Classical correlations are thus obtained in correspondence
of the POVM that minimizes the conditional entropy $\sum_k p_k
S_{V}(\varrho^{\Pi_k}_{A|B})$, allowing one to obtain the highest
amount of information on the state of system $A$. As a matter of fact,
the above definition is in general non symmetric with respect to the
interchange of the subsystems. The quantum discord is then defined as:
\begin{equation}
{\cal D}_{A|B}(\varrho_{AB}) =
I_M(\varrho_{AB}) - {\cal C}_{A|B}(\varrho_{AB})\,,
\end{equation}
$I_M(\varrho_{AB})$ being the mutual information (\ref{mut:info}).
\par
In the particular case of a two-mode Gaussian state, the Gaussian
quantum discord is evaluated addressing only Gaussian measurements
performed on the subsystems and can be written as (for conditional
Gaussian measurements on Gaussian states, see Sect.~\ref{sec:G:meas})
\cite{gio:10,ade:10}:
\begin{subequations}\label{Gdiscord}
\begin{align}
{\cal D}_{A|B}(\varrho_{AB}) 
&= S_{V}(\varrho_{B}) -S_{V}(\varrho_{AB})
+ f\left(\sqrt{E^{\min}_{A|B}}\right)\,,\\
&= f\left(\sqrt{E^{\min}_{A|B}}\right) - S_{A|B}(\varrho_{AB})\,,
\end{align}
\end{subequations}
where $f(x)$ has been defined in Eq.~(\ref{f:x}),
$S_{A|B}(\varrho_{AB})$ is the conditional entropy (\ref{cond12}) and,
in terms of the symplectic invariants (\ref{loc:symp:inv}), $E^{\rm
  min}_{A|B}$ writes \cite{ade:10}:
\begin{align}\label{infdet}
E^{\min}_{A|B} =
\left\{ \begin{array}{ll}
% \frac{{2 I_3^2 - \left(I_1-4I_4\right) \left(I_2-1/4\right)
% + 2 |I_3| \sqrt{I_3^2 -
% \left(I_1-4I_4\right)\left(I_2-1/4\right)}}}
% {4\left(I_2-1/4\right)^2}
\left[
\frac{{|I_3| + \sqrt{I_3^2 -
\left(I_1-4I_4\right)\left(I_2-1/4\right)}}}
{2\left(I_2-1/4\right)}
\right]^2
& \quad\hbox{if }\;
\frac{\left (I_1I_2 - I_4  \right)^2}
{\left(I_1 + 4I_4 \right) \left(I_2 + 1/4 \right) I_3^2} \le 1 \,, \\[3ex]
% \frac{{I_1 I_2 - I_3^2 + I_4-
% \sqrt{I_3^4+\left(I_1 I_2 - I_4 \right)^2 - 2 I_3^2
% \left(I_1 I_2+I_4\right)}}}{{2 I_2}}
\frac{{I_1 I_2  + I_4 - I_3^2 -
\sqrt{\left(I_1 I_2 + I_4 - I_3^2 \right)^2 - 4 I_1I_2I_4
}}}{{2 I_2}}
& \quad\hbox{otherwise.}
\end{array} \right.
\end{align}
In the case of the squeezed thermal state (\ref{c3:thermalsq}),
one has (we set $\xi=r$):
\begin{equation}
\sqrt{E^{\min}_{A|B}} = \frac12 +
\frac{2N_1(1+N_2)}{1-N_1+N_2+(1+N_1+N_2)\,\cosh(2r)}\,,
\end{equation}
and the explicit expression of the quantum discord (\ref{Gdiscord})
can be easily evaluated.
\par
The quantity $f\left(\sqrt{E^{\min}_{A|B}}\right)$ corresponds to the
average von Neumann entropy of the conditional single-mode Gaussian
state in which is left the subsystem $A$ after the Gaussian
measurement on subsystem $B$ minimizing the conditional entropy in
Eq.~(\ref{ClassCorrs}) (see Ref.~\cite{ade:10} for details of the
calculation).  ${\cal D}_{B|A}(\varrho_{AB})$ can be obtained by
exchanging the roles of the two subsystems.
\par
It is worth noting that quantum discord can be nonzero even if the
state is separable, which indicates that entanglement is not the only
source of quantum correlations. For instance, there are examples of
quantum computational algorithms showing a speedup with respect to the
classical counterparts, even in the absence of entanglement
\cite{DatCav,LanWhi}. States with zero discord represent essentially a
classical probability distribution embedded in a quantum system, while
a positive discord, even on separable (mixed) states, is an indicator
of quantumness \cite{terno,dattainprep}, and may operationally be
associated with the impossibility of local broadcasting \cite{piani}.
\par
In the case of pure two-mode Gaussian states, since $\Delta(\bmsigma)
= I_1+I_2+2I_3 = 1/2$ and $I_4=1/16$, one has $S_V(\varrho_{AB}) =
f\left(\sqrt{E^{\min}_{A|B}}\right) = 0$, and hence the Gaussian
quantum discord (\ref{Gdiscord}) reduces to the entropy of
entanglement, i.e., ${\cal D}_{A|B}(\varrho_{AB}) = {\cal
  D}_{B|A}(\varrho_{AB}) = S_V(\varrho_A) = S_V(\varrho_B)$.

%%%%%%%%%%%%%%%%%%%%%%%% SECTION %%%%%%%%%%%%%%%%%%%%%%%%%%
\section{Gaussian states in noisy channels}
\label{sec:GS:noisy}
As one may expect, the dissipative dynamics of a Gaussian states in a
Gaussian environment, or channel, can be reduced to a suitable
transformation of its CM and first-moments vector. In this tutorial we
focus on Markovian environments, however, it is possible to extend the
analysis to non-Markovian ones, as described, for instance, in
Refs.~\cite{hu:92,int:03}.
\par
The dynamics of a single-mode quantum state $\varrho_t$ through a
(Markovian) noisy environment is governed by the following Master
equation:
\begin{equation}\label{c6:me:lind}
\dot{\varrho}_t = \frac{\Gamma}{2}\Big\{
(N+1) \mL [\ha] + N \mL [\ha^\dag]
- M^{*} \mD [\ha] - M \mD [\ha^\dag]
\Big\}\,\varrho_t\,,
\end{equation}
where $\mL [\hO]\varrho_t=2 \hO\varrho_t
\hO^{\dag}-\hO^{\dag}\hO\varrho_t - \varrho_t \hO^{\dag} \hO$ and $\mD
[\hO]\varrho_t = 2 \hO\varrho_t \hO - \hO \hO \varrho_t - \varrho_t
\hO \hO$ are Lindblad superoperators, $\Gamma$ is the overall damping
rate, while $N\in\rr$ and $M\in\cc$ represent the effective number of
photons and the squeezing parameter of the bath, respectively
\cite{FOP:05}.  The terms proportional to $\mL [\ha]$ and to
$\mL [\ha^\dag]$ describe losses and linear, phase-insensitive,
amplification processes, respectively, while the terms proportional to
$\mD [\ha]$ and $\mD [\ha^\dag]$ describe phase dependent
fluctuations. The positivity of the density matrix imposes the
constraint $|M|^2 \le N(N+1)$. At thermal equilibrium, i.e., for
$M=0$, $N$ coincides with the average number of thermal photons in the
bath.
\par
In order to explicitly derive the evolution of the CM and
first-moments vector, we transform the Master equation
(\ref{c6:me:lind}) into the following Fokker-Planck equation for the
Wigner function $W[\varrho_t](\bmX)$ associated with $\varrho_t$
\cite{FOP:05}:
\begin{equation}
\partial_t W[\varrho_t](\bmX) =
\frac{\Gamma}{2} \bigg(\partial_{\bmX}^{\sT} X +
\partial_{\bmX}^{\sT} \bmsigma_{\infty} \partial_{\bmX} \bigg)
W[\varrho_t](\bmX)\label{c6:me:single:cmpct}\,,
\end{equation}
where $\bmX \equiv
(x, y)^{\sT}$, $\partial_{\bmX} \equiv (\partial_x, \partial_y)^{\sT}$, and
we introduced the diffusion matrix  $\bmsigma_\infty$:
\begin{equation}
\bmsigma_{\infty} = \left(\begin{array}{cc}
\left(\frac12 +N\right)+\re{M} & \im{M} \\[1ex]
\im{M} & \left(\frac12 +N\right)-\re{M}
\end{array}\right) \,. \label{c6:diff:infinity}
\end{equation}
\par
If the initial state $\varrho_0$ is a Gaussian state with CM
$\bmsigma_0$ and first-moments vector $\overline{\bmX}_0 \equiv{\rm
  Tr}[\varrho_0 \hat{\bmR}]$, respectively, the Wigner function
$W(\bmX)$ of the evolved state under the action of the
Eq.~(\ref{c6:diff:infinity}) is still Gaussian, but with CM and
first-moments vector given by (see Ref.~\cite{FOP:05} for the explicit
calculation):
\begin{equation}
\bmsigma_t = e^{-\Gamma t} \bmsigma_0 + (1-e^{-\Gamma t})
\bmsigma_{\infty},\quad \mbox{and} \quad
\overline{\bmX}_t = e^{-\Gamma t/2}\,\overline{\bmX}_0,
\end{equation}
respectively, which show that $\bmsigma_{\infty}$ is the asymptotic CM
when the initial state is Gaussian, while $\overline{\bmX}_t
\equiv{\rm Tr}[\varrho_t \hat{\bmR}]$ is damped to zero.
\par
The extension to two-mode or, more in general, to $n$-mode states
interacting with uncorrelated environments, each described by a
Master equation of the form (\ref{c6:me:lind}), is straightforward. In
this case, if $\bmsigma_0$ and $\overline{\bmX}_0$ refer to the CM
and first-moments vector of the initial state $n$-mode state and
$\Gamma_k$, $N_k$ and $M_k$ are the parameter characterizing the
environment interacting with the $k$-th mode, then we have:
\begin{equation}
\bmsigma_t = {\mathbbm G}_t^{1/2} \bmsigma_0 {\mathbbm G}_t^{1/2} +
\left({\mathbbm 1}_{2n}-{\mathbbm G}_t\right)
\bmsigma_{\infty},\quad \mbox{and} \quad
\overline{\bmX}_t = {\mathbbm G}_t^{1/2} \overline{\bmX}_0,
\end{equation}
where ${\mathbbm G}_t = \bigoplus_{h=1}^{n} e^{-\Gamma_h t} \ii_2$ and
$\bmsigma_\infty = \bigoplus_{h=1}^{n} \bmsigma_{h,\infty}$ with:
\begin{equation}
\bmsigma_{h,\infty} = \left(\begin{array}{cc}
\left(\frac12 +N_{h}\right)+\re{M_{h}} & \im{M_{h}} \\ [1ex]
\im{M_{h}} & \left(\frac12 +N_{h}\right)-\re{M_{h}}
\end{array}\right) \,. \label{c6:canalino}
\end{equation}
\par
Starting form $\bmsigma_{t}$ and $\overline{\bmX}_{t}$, one can easily
evaluate the evolution of all the quantities addressed in the previous
sections, such as purity and, for two-mode states, the separability
thresholds and entropies (the interested reader can find the explicit
calculations and a thorough analysis, e.g., in Ref.~\cite{FOP:05} and
references therein).

%%%%%%%%%%%%%%%%%%%%%%%% SECTION %%%%%%%%%%%%%%%%%%%%%%%%%%
\section{Gaussian measurements onto a Gaussian state}
\label{sec:G:meas}
In the previous sections we have reviewed how a Gaussian state can be
generated and characterized. We have also addressed its kinematics and
evolution through noisy channels. In order to make this tutorial as
complete as possible, we now focus on conditional Gaussian
measurements \cite{gie:02,eis:03,oli:EPJST,sas:08}, such as homodyne
detection and double homodyne detection \cite{FOP:05}, performed on
Gaussian states (for a PhD tutorial on the manipulation of Gaussian
states at the photon level see, e.g.,
Ref.~\cite{kim:JPB}). Conditional measurements are extremely important
in quantum information processing, since they are at the basis of
quantum teleportation and telecloning protocols and allow to generate
and manipulate new classes of states \cite{par:JOB:03}. Furthermore, a
single homodyne detector has been recently used to fully characterize
a two-mode squeezed thermal state \cite{dau:09,buo:10}.
\par
In order to show how a typical calculation involving Gaussian states
and operations is carried out, we explicitly derive the characteristic
function of a conditional Gaussian state obtained by performing a
Gaussian measurement on one of its $n$ modes.
\par
Let us consider the following Gaussian characteristic function with
zero first-moments vector (extension to non-zero first-moments
states is straightforward) associated with a $n$-mode state $\varrho$
[for the sake of simplicity we use the characteristic function as
defined in Eq.~(\ref{chi:simple}) and drop the explicit dependence on
the operators]:
\begin{equation}\label{CF}
\chi(\bmLambda) = \exp\left\{
-\mbox{$\frac12$} \bmLambda^{T} \bmSigma\, \bmLambda
\right\}\,,
\end{equation}
where $\bmLambda = (\bmLambda_1, \bmLambda_2, \dots, \bmLambda_n)^{\sT}
\in {\mathbbm R}^{2n}$ is a column vector and $\bmSigma$ is the $2n
\times 2n$ CM. Now we assume to perform a Gaussian measurement on one
of the modes, that is a measurement described by a POVM with Gaussian
characteristic function. Without lack of generality, we can assume
that the measurement involves mode 1, and, thus, the corresponding
characteristic function may be written as:
\begin{equation}\label{POVM:1}
\chi_{\rm M}(\bmLambda_1) = \pi^{-1}\,\exp\left\{
-\mbox{$\frac12$} \bmLambda_1^{T} \bmsigma_{\rm M}\, \bmLambda_1
- i \bmLambda_1^{T}\bmX \right\}\,,
\end{equation}
where $\bmsigma_{\rm M}$ and $\bmX$ are the CM the first-moments
vector or, more precisely, the outcome of the measurement,
respectively. For the sake of simplicity, we write $\bmLambda$ and the
CM $\bmSigma$ in the following block form:
\begin{equation}
\bmLambda = (\bmLambda_1, \bmLambda_2, \dots, \bmLambda_n)^{\sT} =
(\bmLambda_1,\tilde{\bmLambda})^{\sT}\,, \quad \mbox{and}\quad
\bmSigma = \left(
\begin{array}{cc}
\bmA & \bmC \\
\bmC^{T} & \bmB
\end{array}
\right)\,,
\end{equation}
where $\bmA \in {\mathbbm R}^{2}\times {\mathbbm R}^{2}$ and $\bmB \in
{\mathbbm R}^{2(n-1)}\times {\mathbbm R}^{2(n-1)}$ are symmetric, and
$\bmC \in {\mathbbm R}^{2}\times {\mathbbm R}^{2(n-1)}$, making
evident the mode undergoing the measurement. The conditional
characteristic function of the system after the measurement with
outcome $\bmX$ is:
\begin{equation}\label{integral}
\chi'(\tilde{\bmLambda}) = 
\frac{1}{p(\bmX)}
\int_{{\mathbbm R}^2}\frac{d^2\bmLambda_{1}}{2\pi}
\chi(\bmLambda_1,\tilde{\bmLambda})\,
\chi_{\rm M}(-\bmLambda_1)\,,
\end{equation}
where we used the trace rule (\ref{Chi:trace}) and $p(\bmX)$ is the
probability of the outcome $\bmX$:
\begin{align}
p(\bmX) &= \int_{{\mathbbm R}^{2n}}
\frac{d^2\bmLambda_{1}\, d^{2(n-1)}\tilde{\bmLambda}}{(2\pi)^{n}}\,
\chi(\bmLambda_1,\tilde{\bmLambda})
\,\chi_{\rm M}(-\bmLambda_1)\, (2\pi)^{(n-1)}\delta(-\tilde{\bmLambda})\\
&= \frac{ \exp\left\{
-\mbox{$\frac12$} \bmX^{T} (\bmA + \bmsigma_{\rm M})^{-1}\,
\bmX \right\}
}{\pi\sqrt{\det[\bmA + \bmsigma_{\rm M}]}}\,,\label{p:out:1}
\end{align}
where $\delta(-\tilde{\bmLambda}) =
\prod_{k=2}^{n} \delta^{(2)}(-\bmLambda_k)$ is the product of Kronecker deltas
in ${\mathbbm R}^{2}$. Note that:
\begin{align}
\chi(\bmLambda_1,\tilde{\bmLambda})\,
\chi_{\rm M}(-\bmLambda_1) = \pi^{-1}\,\exp\left\{
-\mbox{$\frac12$} (\bmLambda_1,\tilde{\bmLambda})^{T}
\bmsigma \, (\bmLambda_1,\tilde{\bmLambda})
+ i \bmLambda_1^{T}\bmX \right\}\,,
\end{align}
with:
\begin{equation}
\bmsigma = \left(
\begin{array}{cc}
\bmA + \bmsigma_{\rm M} & \bmC \\
\bmC^{T} & \bmB
\end{array}
\right)\,.
\end{equation}
In order to perform the integral (\ref{integral}) we observe that
$\bmsigma$ can be rewritten as follows:
\begin{equation}
\bmsigma = \bmM^{T} \left(
\begin{array}{cc}
\bmA + \bmsigma_{\rm M} & \boldsymbol 0 \\
\boldsymbol 0 & \bmB-\bmC^{T}(\bmA + \bmsigma_{\rm M} )^{-1}\bmC
\end{array}
\right) \bmM\,, \quad
\bmM = \left(
\begin{array}{cc}
{\mathbbm 1}_{2} & (\bmA + \bmsigma_{\rm M})^{-1}\bmC \\
{\boldsymbol 0} & {\mathbbm 1}_{2(n-1)}
\end{array}
\right)\,.
\end{equation}
The matrix $\bmB-\bmC^{T}(\bmA + \bmsigma_{\rm M})^{-1}\bmC$ is the
Schur complement of the matrix $\bmsigma$ with respect to $\bmA
+ \bmsigma_{\rm M}$. Now, since:
\begin{equation}
\bmM (\bmLambda_1,\tilde{\bmLambda}) =
\Big(\bmLambda_1 + (\bmA + \bmsigma_{\rm M})^{-1}\bmC \tilde{\bmLambda},
\tilde{\bmLambda}\Big)\,,
\end{equation}
Eq.~(\ref{integral}) reduces to:
\begin{align}
\chi'(\tilde{\bmLambda}) &= \frac{1}{p(\bmX)}\,
\exp\left\{ -\mbox{$\frac12$} \tilde{\bmLambda}^{T}
[\bmB-\bmC^{T}(\bmA + \bmsigma_{\rm M})^{-1}\bmC]\,
\tilde{\bmLambda}
- i \tilde{\bmLambda}^{T}\bmC^{T}(\bmA + \bmsigma_{\rm M})^{-1}\bmX
\right\}\nonumber\\
&\hspace{0.5cm}\times
\int_{{\mathbbm R}^2}\frac{d^2\bmLambda'}{2\pi^{2}}
\exp\left\{ -\mbox{$\frac12$}
(\bmLambda')^{T}
(\bmA + \bmsigma_{\rm M})\,
(\bmLambda') - i (\bmLambda')^{T}\bmX\right\}\,,\\
&=
\exp\left\{ -\mbox{$\frac12$} \tilde{\bmLambda}^{T}
[\bmB-\bmC^{T}(\bmA + \bmsigma_{\rm M})^{-1}\bmC]\, \tilde{\bmLambda}
-i \tilde{\bmLambda}^{T}\bmC^{T}(\bmA + \bmsigma_{\rm M})^{-1} \bmX
\right\}\,,\label{out:1}
\end{align}
where we performed the change of variables $\bmLambda' = \bmLambda_1 +
(\bmA + \bmsigma_{\rm M})^{-1}\bmC \tilde{\bmLambda}$.  The
conditional state $\chi'(\tilde{\bmLambda})$ is a $(n-1)$-mode
Gaussian state with CM $\bmB-\bmC^{T}(\bmA + \bmsigma_{\rm
  M})^{-1}\bmC$ and first-moments vector $\bmC^{T}(\bmA +
\bmsigma_{\rm M})^{-1} \bmX$.  Analogously, if we carry out the
measurement on the mode $n$ (actually, the last one), we obtain that
the conditional state is still Gaussian but with CM $\bmA-\bmC (\bmB +
\bmsigma_{\rm M})^{-1}\bmC^{T}$ and first-moments vector $\bmX^T (\bmB
+ \bmsigma_{\rm M})^{-1} \bmC$, where, now, $\bmA \in {\mathbbm
  R}^{2(n-1)}\times {\mathbbm R}^{2(n-1)}$, $\bmB \in {\mathbbm
  R}^{2}\times {\mathbbm R}^{2}$ and $\bmC \in {\mathbbm
  R}^{2(n-1)}\times {\mathbbm R}^{2}$.

%%%%%%%%%%%%%%%%%%%%%%%% SECTION %%%%%%%%%%%%%%%%%%%%%%%%%%
\section{Fidelity between Gaussian states}
\label{sec:fidelity}
The fidelity is one of the most important figure of merit in quantum
information and quantifies the similarity between two states
$\varrho_1$ and $\varrho_2$. The Uhlmann's fidelity is defined as
\cite{uhlm:76}:
\begin{equation}\label{fid:12}
{\cal F}(\varrho_1,\varrho_2) =
\left\{\hbox{Tr}\left[(\sqrt{\varrho_1}\varrho_2\sqrt{\varrho_1})^{1/2}\right]\right\}^2,
\end{equation}
and corresponds to the maximal transition probability between all
purifications of the two states.
\par
In the case of single-mode Gaussian states with CMs $\bmsigma_k$
and first-moments vectors $\overline{\bmX}_k$, $k=1,2$,
Eq.~(\ref{fid:12}) leads to \cite{scu:98}:
\begin{equation}
{\cal F}(\varrho_1,\varrho_2) =
\frac{\exp\left\{ -\mbox{$\frac12$}\,
(\overline{\bmX}_{1}-\overline{\bmX}_{2})^{T} (\bmsigma_1+\bmsigma_2)^{-1}
(\overline{\bmX}_{1}-\overline{\bmX}_{2}) \right\}}{
  \sqrt{\Delta +\delta} - \sqrt{\delta}},
\end{equation}
with $\Delta =
\det[\bmsigma_1 + \bmsigma_2]$ and $\delta = 4\prod_{k=1}^{2}
(\det[\bmsigma_k]-\frac14)$.
\par
The problem of finding an analytical formula for the fidelity between
$n$-mode Gaussian states has been very recently solved in an elegant
way \cite{mm:11}. In particular, for two-mode Gaussian states one
obtains:
\begin{equation}\label{two:fid}
{\cal F}(\varrho_1,\varrho_2) = \hbox{Tr}[\varrho_1\varrho_2]\,
\left( \sqrt{{\cal X}} + \sqrt{{\cal X}-1}\right)^2,
\end{equation} 
where ${\cal X} = 2\sqrt{{\cal A}} + 2\sqrt{{\cal B}} + \frac12$ and:
\begin{align}
{\cal A} =
\frac{\det[\bmOmega\,\bmsigma_1\,\bmOmega\,\bmsigma_2-\frac14
  \mathbbm{1}_{4}]}{\det[\bmsigma_1+\bmsigma_2]}\,, \quad
{\cal B} =
\frac{\det[\bmsigma_1+\frac{i}{2}\bmOmega]\,
\det[\bmsigma_2+\frac{i}{2}\bmOmega]}{\det[\bmsigma_1+\bmsigma_2]}\ge 0\,,
\end{align}
are symplectic invariants, $\bmOmega =
\boldsymbol{\omega}\oplus\boldsymbol{\omega}$ is the symplectic matrix
(\ref{defOM}) and, using Eq.~(\ref{Wtrace}) or Eq.~(\ref{Chi:trace}),
we have:
\begin{equation}
\hbox{Tr}[\varrho_1\varrho_2]=
\frac{\exp\left\{ -\mbox{$\frac12$}\,
(\overline{\bmX}_{1}-\overline{\bmX}_{2})^{T} (\bmsigma_1+\bmsigma_2)^{-1}
(\overline{\bmX}_{1}-\overline{\bmX}_{2})\right\}}
{\sqrt{\det[\bmsigma_1+\bmsigma_2]}}.
\end{equation}
The reader can find the full analysis leading to Eq.~(\ref{two:fid})
and the extension to $n$-mode Gaussian states in the original paper
\cite{mm:11}.

%%%%%%%%%%%%%%%%%%%%%%%% SECTION %%%%%%%%%%%%%%%%%%%%%%%%%%
\section{Conclusions}
\label{sec:remarks}
In this tutorial we have presented the basic tools and results in
order to deal with Gaussian states in phase space. We have seen how
their generation, manipulation and propagation through noisy channels
can be described by means of suitable transformations of the
covariance matrix and first-moments vector. Focusing on two-mode
Gaussian states, we have addressed their characterization by means of
mutual information and conditional entropies, their separability and
entanglement properties. As a pedagogical example, we have explicitly
showed how to calculate the conditional state of a multimode Gaussian
state that has gone through a Gaussian measurement on one of its
modes. We have also presented the latest results about the fidelity
between Gaussian states, that is the most important figure of merit in
quantum information.
\par
After the last meeting I had with prof.~Federico Casagrande, I
promised him we would have discussed my research on Gaussian states
and my latest results. Unfortunately, we had not enough time. Of
course, these pages cannot substitute an afternoon spent with him, his
curiosity and his enthusiasm$\ldots$ Nevertheless, I believe he would
have appreciated this tutorial and I hope it could be a useful tool
for students and scientists interested in quantum optics, the main
topic investigated by Federico during his activity at the University
of Milano.

%%%%%%%%%%%%%%%%%%%% ACKNOWLEDGMENTS
\begin{acknowledgement}
  I'd like to thank M.~G.~A.~Paris, F.~Benatti and P.~Marian for
  useful suggestions and discussions. Financial support from the
  University of Trieste (FRA 2009) and MIUR (FIRB RBFR10YQ3H) is
  acknowledged.
\end{acknowledgement}

\end{document}